%% file: SLPRAM-arXiV.tex
\documentclass[11pt]{article}  

\usepackage{fullpage}
\usepackage{epsfig}
\usepackage{graphics}
\usepackage{amsmath,amssymb}
\usepackage{multirow}
\usepackage{cite}
\usepackage{hyperref}

\newcommand{\size}{\ensuremath{\mathrm{size}}}

\newcommand{\heavy}{\ensuremath{\mathrm{heavy}}}
\newcommand{\weight}{\ensuremath{\mathrm{weight}}}

\newcommand{\occ}{\ensuremath{\mathrm{occ}}}

\newcommand{\Rank}{\ensuremath{\mathrm{rank}}}
\newcommand{\Select}{\ensuremath{\mathrm{select}}}
\newcommand{\excess}{\ensuremath{\mathrm{excess}}}
\newcommand{\exsum}{\ensuremath{\mathrm{sum}}}
\newcommand{\fwd}{\ensuremath{\mathrm{fwd\_search}}}
\newcommand{\bwd}{\ensuremath{\mathrm{bwd\_search}}}
\newcommand{\excmin}{\ensuremath{\mathrm{m}}}
\newcommand{\excmax}{\ensuremath{\mathrm{M}}}
\newcommand{\minrmq}{\ensuremath{\mathrm{rmq}}}
\newcommand{\minrmqi}{\ensuremath{\mathrm{rmqi}}}
\newcommand{\maxrmq}{\ensuremath{\mathrm{RMQ}}}
\newcommand{\maxrmqi}{\ensuremath{\mathrm{RMQi}}}

\newcommand\nodex{\ensuremath{x}}
\newcommand\nodey{\ensuremath{y}}

\newcommand\fchild{\ensuremath{\mathtt{first\_child}}}
\newcommand\lchild{\ensuremath{\mathtt{last\_child}}}
\newcommand\parent{\ensuremath{\mathtt{parent}}}
\newcommand\nsibling{\ensuremath{\mathtt{next\_sibling}}}
\newcommand\psibling{\ensuremath{\mathtt{prev\_sibling}}}

\newcommand\desc{\ensuremath{\mathtt{desc}}}

\newcommand\height{\ensuremath{\mathtt{height}}}
\newcommand\LCA{\ensuremath{\mathtt{LCA}}}

\newcommand\levelAnc{\ensuremath{\mathtt{level\_anc}}}
\newcommand\levelLeft{\ensuremath{\mathtt{level\_left}}}
\newcommand\levelRight{\ensuremath{\mathtt{level\_right}}}
\newcommand\levelPred{\ensuremath{\mathtt{level\_pred}}}
\newcommand\levelSucc{\ensuremath{\mathtt{level\_succ}}}
\newcommand\depth{\ensuremath{\mathtt{depth}}}

\newcommand\leftLeaf{\ensuremath{\mathtt{left\_leaf}}}
\newcommand\rightLeaf{\ensuremath{\mathtt{right\_leaf}}}
\newcommand\post{\tiny \textsc{POST}}
\newcommand\pre{\tiny \textsc{PRE}}

\newtheorem{lemma}{Lemma}
\newtheorem{theorem}{Theorem}

\newenvironment{proof}{\noindent\emph{Proof. }}

\begin{document}

\title{Random Access to Grammar-Compressed Strings and Trees\thanks{A preliminary version of this paper appeared in the Proceedings of the 22nd Annual ACM-SIAM Symposium on Discrete Algorithms, 2011.}}

\author{Philip Bille \thanks{DTU Informatics, Technical University of Denmark, Denmark. \href{mailto:phbi@imm.dtu.dk}{phbi@imm.dtu.dk}}
\and 
Gad M. Landau  \thanks{Department of Computer  Science, University of Haifa, Israel. \href{mailto:landau@cs.haifa.ac.il}{landau@cs.haifa.ac.il}}
\and 
Rajeev Raman  \thanks{Department of Computer Science, University of Leicester, UK. \href{mailto:r.raman@mcs.le.ac.uk}{r.raman@mcs.le.ac.uk}. Raman was supported by Royal Society Travel Grant TG091629.} 
\and 
Kunihiko Sadakane \thanks{National Institute of Informatics, Japan. \href{mailto:sada@nii.ac.jp}{sada@nii.ac.jp}. Sadakane was supported in part by JSPS KAKENHI 23240002.}
\and 
Srinivasa Rao Satti \thanks{School of Computer Science and Engineering, Seoul National University, S. Korea. \href{mailto:ssrao@cse.snu.ac.kr}{ssrao@cse.snu.ac.kr}. Satti was partially supported by National Research Foundation of Korea (grant number 2012-0008241).}
\and 
Oren Weimann \thanks{Department of Computer  Science, University of Haifa, Israel. \href{mailto:oren@cs.haifa.ac.il}{oren@cs.haifa.ac.il}. Weimann was partially supported by the Israel Science Foundation grant 794/13.}
}

\date{}
\maketitle

\begin{abstract}
 Grammar based compression, where one replaces a long
string by a small context-free grammar that generates the string, is
  a simple and powerful paradigm that captures many of
  the popular compression schemes, including the Lempel-Ziv family,
 Run-Length Encoding, Byte-Pair Encoding, Sequitur, and Re-Pair. 
 In this paper, we present a novel grammar representation that allows efficient random access to any character or substring  without decompressing the string.  
 
Let $S$ be a string of length $N$ compressed into a context-free grammar $\mathcal{S}$ of size $n$. We present two representations of $\mathcal{S}$  achieving $O(\log N)$ random access time, and either $O(n\cdot  \alpha_k(n))$ construction time and space on the pointer machine model, or $O(n)$ construction time and space 
on the RAM. Here, $\alpha_k(n)$ is the inverse of the $k^{th}$ row of Ackermann's function. Our representations also efficiently support decompression of any substring in $S$:  we can  decompress any substring of length $m$ in the same complexity as a single random access query and additional $O(m)$ time.  Combining these results with fast
  algorithms for uncompressed approximate string matching
  leads to several efficient algorithms for approximate string matching on grammar-compressed strings without decompression.  For instance, we can find all
  approximate occurrences of a pattern $P$ with at most $k$
  errors in time $O(n(\min\{|P|k, k^4 + |P|\} + \log N) + \occ)$, where
  $\occ$ is the number of occurrences of $P$ in $S$. Finally, we generalize our results to navigation and other operations on 
grammar-compressed ordered trees.

All of the above bounds significantly improve the currently best known
results. To achieve these bounds, we introduce several new techniques
and data structures of independent interest, including a predecessor data structure, two ``biased" weighted ancestor data structures, and a compact representation of heavy paths in grammars. 
\end{abstract}

\input{Introduction}

\input{LinearSpace}

\input{IntervalBiased}

\input{Tradeoffs}

\input{biasedskip}

\input{comp-trees}

\input{conclusions}

\bibliographystyle{abbrv} 
\bibliography{SLPRAM} 

\end{document}

%% file: Introduction.tex
\section{Introduction}
Modern textual or semi-structured databases, e.g. for biological and WWW data,
are huge, and are typically stored in compressed
form. A query to such databases will typically
retrieve only a small portion of the data.
This presents several challenges: how to query the
compressed data directly and efficiently, without the
need for additional data structures (which can be
many times larger than the compressed data), and how to retrieve the
answers to the queries.  In many practical cases, the naive approach
of first decompressing the entire data and then processing it
is completely unacceptable -- for instance XML data
compresses by an order of magnitude on disk \cite{FerraginaLMM09}
but \emph{expands} by an order of magnitude when represented
in-memory \cite{DelprattRR08}; as we will shortly see, this approach
is very problematic from an asymptotic perspective as well.  
Instead we want to support this functionality directly on the compressed data. 

We focus on two data types, \emph{strings} and \emph{ordered trees}, and
consider the former first. Let $S$ be a string of length $N$ from
an alphabet $\Sigma$, given in a compressed representation $\mathcal{S}$ of size $n$.
The \emph{random access problem} is to compactly represent $\mathcal{S}$
while supporting fast random access queries, that is, given an index
$i$, $1\leq i\leq N$, report $S[i]$. More generally, we want to
support \emph{substring decompression}, that is, given a pair of
indices $i$ and $j$, $1\leq i \leq j\leq N$, report the substring
$S[i]\cdots S[j]$. The goal is to use little space for the
representation of $\mathcal{S}$ while supporting fast random access
and substring decompression. Once we obtain an efficient substring decompression method, 
it can also serve as a basis for a
compressed version of classical pattern matching. 
For example, given an (uncompressed) pattern
string $P$ and $\mathcal{S}$, the \emph{compressed pattern matching
problem} 
is to find all occurrences of $P$ within $S$ 
more efficiently than to naively decompress $\mathcal{S}$ into $S$ and 
then search for $P$ in $S$. 
An important variant of the pattern matching
problem is when we allow approximate matching (i.e., when $P$ is 
allowed to appear in $S$ with some errors).  


\begin{figure}[h!]
\begin{center}
\includegraphics[scale=0.4]{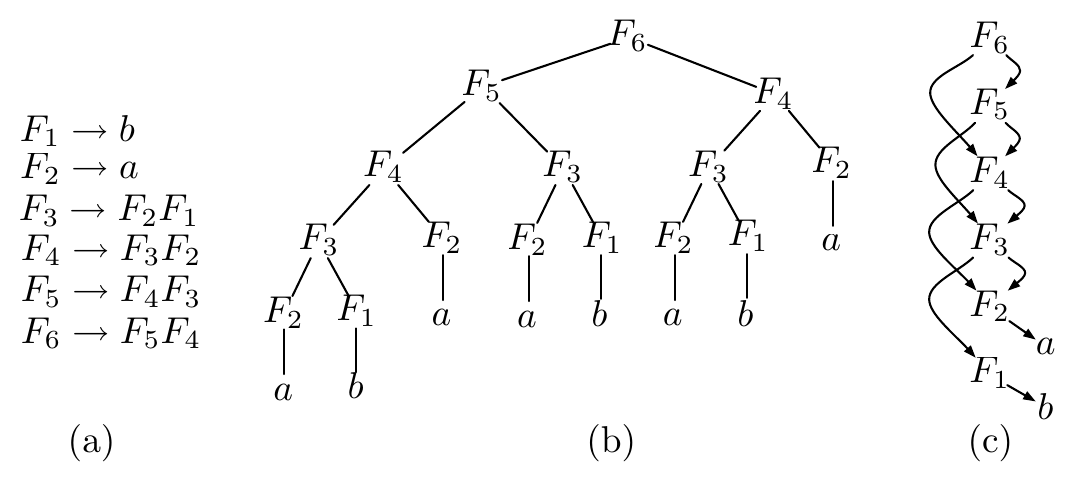} 
\caption{(a) A context-free grammar generating the string abaababa. (b)
  The corresponding parse tree. (c) The acyclic graph defined by the
  grammar.}
\label{grammar}
\end{center}
\end{figure}
We consider these problems in the context of \emph{grammar-based
compression}, where one replaces a long string by a small
context-free grammar (CFG) that generates this string (and this string only. We assume
without loss of generality that the grammars are in fact
\emph{straight-line programs} (SLPs) and so on the righthand side of
each grammar rule there are either exactly two variables or one terminal
symbol, see
Fig.~\ref{grammar}(a)). Such grammars capture many 
popular compression schemes including the
Lempel-Ziv family~\cite{ZivLempel1976, ZivLempel1977, Welch1984},
Sequitur~\cite{Nevill-ManningWitten1997}, Run-Length Encoding,
Re-Pair~\cite{LarssonMoffat2000}, and many
more~\cite{Shibata-et-al-1999,Gage1994,
KiefferYang2000,Kiefferetal2000,YangKieffer2000,ApostolicoLonardi2000,ApostolicoLonardi1998,ApostolicoLonardi2000a}. All of these are or can be transformed into equivalent grammar-based
compression schemes with little 
expansion~\cite{Rytter2003,Charikaretal2005}. In general, the size of
the grammar, defined as the total number of symbols in all derivation
rules, can be exponentially smaller than the string it generates.
From an algorithmic perspective, the properties of compressed data were
used to accelerate the solutions to classical problems on
strings including exact pattern
matching~\cite{AmirLandauSokol2003,KarkkainenUkkonen1996,Lifshits2007,Manber1994,Shibata-et-al-2000}
and approximate pattern
matching~\cite{AmirBensonFarach1996,ArbellLandauMitchell2001,
  BunkeCsirik1995,CegielskiGuessarianLifshitsMatiyasevich2006,CrochemoreLandauZiv-Ukelson2003,HermelinLandauLandauWeimann2009,BilleFagerbergGoertz2007,KarkkainenNavarroUkkonen2000,KarkkainenUkkonen1996,MakinenNavarroUkkonen1999,NavarroKidaetal2001}.

We also consider the problem of representing an ordered
rooted tree $T$ (of arbitrary degree) with $N$ nodes.  
We assume that $T$ is represented as a \emph{balanced parenthesis} 
sequence \cite{MunroRaman01}, which is obtained by traversing $T$ in pre-order 
and outputting `(' upon entering a node for the first time, and `)' upon 
leaving it. We assume that this balanced
parenthesis sequence is given as an SLP of size $n$, and consider the problem
of performing operations on $T$.  

This compression method may seem a little artificial, but it is in fact a powerful technique that captures existing tree compression methods.  For example, a popular tree
compression method is to represent it by the minimal DAG obtained by sharing identical subtrees,
giving a DAG with $e$ edges (see Fig.~\ref{fig:compressedtree} in Section~\ref{sec:labelledtrees}).
We note that this DAG can be represented as an SLP of size $O(e)$ that generates the BP sequence
of the tree.  However, the SLP representation of the BP string can be much smaller than the 
minimal DAG: for example, if $T$ is just a line with $N$ nodes, 
then DAG compression is completely ineffective, but the BP sequence of $T$, namely
$\mbox{\rm (}^N \mbox{\rm )}^N$, is generated by an SLP of size $O(\log N)$.

\subsubsection*{Our Results}

We present new representations of grammar compressed
strings and trees. We consider two
models, the \emph{pointer machine} \cite{Tarjan83} and the \emph{word RAM} (henceforth
just RAM) \cite{Hagerup98}. We further make the assumption that all memory cells 
can contain $\log N$-bit integers -- this many bits are needed 
just to represent the input to a random access query. 
Let $\alpha_k(n)$ be the inverse of the $k^{th}$ row of Ackermann's
function\footnote{The inverse Ackermann function $\alpha_k(n)$ can be
  defined by $\alpha_k(n)=1+\alpha_k(\alpha_{k-1}(n))$ so that
  $\alpha_1(n)= n/2$, $\alpha_2(n)=\log n$, $\alpha_3(n)=\log^* n$,
  $\alpha_4(n)=\log^{**} n$ and so on.  Here, $\log^{**} n$ is the
  number of times the $\log^*$ function is applied to $n$ to produce a
  constant.}. For strings, we show:
%
%

\begin{theorem}\label{thm:substringdecompression}
For an SLP $\mathcal{S}$ of size $n$ representing a string of length
$N$ we can decompress a substring of length $m$ in time $O(m + \log N)$
\begin{itemize}
\item[(i)] after $O(n\cdot \alpha_k(n) )$
  preprocessing time and space for any fixed $k$, or,
\item[(ii)] after $O(n)$ preprocessing
  time and space on the RAM model. 
\end{itemize}
\end{theorem}

\noindent Next, we show how to combine
Theorem~\ref{thm:substringdecompression} with any black-box
(uncompressed) approximate string matching algorithm to solve the corresponding
compressed approximate string matching problem over grammar-compressed strings.
We obtain the following connection between classical (uncompressed)
and grammar compressed approximate string matching. Let $t(m)$ and $s(m)$ be the
time and space bounds of some (uncompressed) approximate string matching
algorithm on strings of lengths $O(m)$, and let $\occ$ be the number
of occurrences of $P$ in $S$.

\begin{theorem}\label{thm:approxstringmatching}
Given an SLP $\mathcal{S}$ of size $n$ representing a string of length
$N$ and a string $P$ of length $m$ we can find all approximate
occurrences of $P$ in $\mathcal{S}$ in time $O(n(m + t(m) + \log N) + \occ)$ and
\begin{itemize}
\item[(i)] in space $O(n\cdot \alpha_k(n) + m + s(m))$ on the pointer machine model and
\item[(ii)] in space $O(n + m + s(m) +  \occ)$ on the RAM model. 
\end{itemize}
\end{theorem}

Coming to the tree representation problem, suppose that nodes of
the uncompressed tree $T$ are numbered $1,\ldots,N$ in pre-order, and
$T$ is represented as an SLP 
that generates its BP sequence. We are mainly concerned with navigation operations in the tree 
such as $\mbox{\rm parent}(i)$ and $\mbox{\rm lca}(i,j)$, which return the 
(pre-order) numbers of the node 
that is the parent of $i$ or the LCA of $i$ and $j$, respectively
(a full list of navigation operations can be found in 
Table~\ref{tab:navops}).  We show:
\begin{theorem}\label{thm:treerep}
Given an SLP of size $n$ that represents the BP sequence of
a rooted ordered tree $T$ with $N$ nodes, we can support the 
navigation operations given in Table~\ref{tab:navops} in $O(\log N)$ time using:
\begin{itemize}
\item[(i)] $O(n \alpha_k(n))$ words and preprocessing time on the pointer machine model;
\item[(ii)] $O(n)$ words and preprocessing time on the RAM model.
\end{itemize}
\end{theorem}

\noindent
\emph{Remark:} In the discussed applications above, it is more appropriate to 
consider \emph{labelled} trees \cite{FerraginaLMM09}, where 
each node is labelled with a character from some alphabet $\Sigma$. 
A basic operation on the labels is $\mbox{\rm access}(i)$, which returns the 
symbol associated with node $i$. This can
be readily implemented in $O(\log N)$ time by SLP-compressing the string that
comprises the labels of $T$ in pre-order, and using 
Theorem~\ref{thm:substringdecompression}.
Note that separately SLP-compressing the tree structure and the 
labels of $T$ in pre-order cannot be asympotically worse than 
SLP-compressing (say) a 
``labelled" parenthesis string, obtained by outputting 
`$\mbox{(}_c$' upon entering a node labelled $c$, and `$\mbox{)}_c$' upon 
leaving it.

\begin{table}
\label{tab:navops}
\caption{Navigational Operations on an Ordered Tree.}
\begin{center}
\begin{tabular}{rl}
$\parent(\nodex)$ & parent of node $\nodex$\\

$\fchild(\nodex)$ & first child of node $\nodex$\\

$\lchild(\nodex)$ & last child of node $\nodex$\\

$\nsibling(\nodex)$ & next (previous) sibling of node $\nodex$\\

$(\psibling(\nodex)$) & \\

$\depth(\nodex)$ & depth of node $\nodex$\\

$\levelAnc(\nodex,i)$ & ancestor of node $\nodex$ that is $i$ levels above $\nodex$, for $i\geq0$\\

$\desc(\nodex)$ & number of descendants (subtree size) of node $\nodex$\\

$\height(\nodex)$ & returns the height of the subtree rooted at node $\nodex$\\

$\LCA(\nodex, \nodey)$ & returns the lowest common ancestor of the nodes $\nodex$ and $\nodey$\\

$\leftLeaf(\nodex)$ & leftmost (rightmost) leaf of the subtree rooted at node $\nodex$\\
($\rightLeaf(\nodex)$) & \\

$\Rank_{\pre/\post}(\nodex)$ & position of $\nodex$ in the preorder or postorder traversal of the tree\\

$\Select_{\pre/\post}(j)$ & $j$-th node in the preorder or postorder traversal of the tree\\

$\levelLeft(i)$ & first (last) node visited in a preorder traversal among all the\\ 
($\levelRight(i)$) & $\quad$ nodes whose depths are $i$\\

$\levelSucc(\nodex)$ & {\em level successor} ({\em predecessor)} of node $\nodex$, i.e. the node visited\\ 
($\levelPred(\nodex)$) & $\quad$ immediately after (before) node $\nodex$ in a preorder traversal \\
& $\quad$ among all the nodes that are at the same level as node $\nodex$.
\end{tabular}
\end{center}
\end{table}
\subsection*{Related Work}

We now describe how our work relates to existing results. 

\subsubsection*{The random access problem} 
If we use $O(N)$ space we can access
any character in constant time by storing $S$ explicitly in an
array. Alternatively, 
we can compute and store the sizes of strings derived by
each grammar symbol in $\mathcal{S}$. This only requires $O(n)$ space
and allows to simulate a top-down search expanding the grammar's
derivation tree in constant time per node. Consequently, a random
access takes time $O(h)$, where $h$ is the height of the derivation
tree and can be as large as $\Omega(n)$. Although any SLP of size $n$ generating
a string of length $N$ can be converted into an SLP 
with derivation tree height $O(\log N)$ \cite{Rytter2003,Charikaretal2005},
the size of the SLP increases to $O(n \log N)$.
Thus, the simple top-down traversal either has poor worst-case
performance or uses non-linear space. Surprisingly, 
the only known improvement to the simple top-down traversal is
a recent succinct representation of
grammars, due to Claude and Navarro~\cite{ClaudeNavarro2009}. They
reduce the space from $O(n\log N)$ \emph{bits} to $O(n \log n) + n\log
N$ bits at the cost of increasing the query time to $O(h \log n)$.

\subsubsection*{The substring decompression problem}
Using the simple random access trade-off we get an $O(n)$ space
solution that supports substring decompression in $O(hm)$
time. Gasieniec et al.~\cite{Gasieniecetal2005, GasieniecPopatov2003}
showed how to improve the decompression time to $O(h + m)$ while
maintaining $O(n)$ space. Also, the representation 
of~\cite{ClaudeNavarro2009} supports
substring decompression in time $O((h+m) \log n)$.

\subsubsection*{The compressed pattern matching problem} 
In approximate pattern matching, we are given two strings
$P$ and $S$ and an \emph{error threshold} $k$.  The goal is to
find all ending positions of substrings of $S$ that are ``within
distance $k$'' of $P$ under some metric, e.g. the 
\emph{edit distance} metric, where the distance is the 
number of edit operations 
needed to convert one substring to the other.

In classical (uncompressed) approximate pattern matching,
a simple algorithm \cite{Sellers1980} solves this
problem (under edit distance) in $O(Nm)$ time and $O(m)$ space, where
$N$ and $m$ are the lengths of $S$ and $P$ respectively. Several
improvements of this result are known (see e.g. \cite{Navarro2001a}). 
Two well-known improvements for small
values of $k$ are the $O(Nk)$ time algorithm of Landau and
Vishkin~\cite{LV1989} and the $O(Nk^4/m + N)$ time algorithm of Cole
and Hariharan~\cite{CH2002}. Both of these can be implemented in
$O(m)$ space.
The use of compression led to many speedups using various compression
schemes~\cite{AmirBensonFarach1996,ArbellLandauMitchell2001,
  BunkeCsirik1995,CegielskiGuessarianLifshitsMatiyasevich2006,CrochemoreLandauZiv-Ukelson2003,HermelinLandauLandauWeimann2009,BilleFagerbergGoertz2007,KarkkainenNavarroUkkonen2000,KarkkainenUkkonen1996,MakinenNavarroUkkonen1999,NavarroKidaetal2001}. 
The most closely
related to our work is approximate pattern matching for LZ78 and LZW
compressed strings~\cite{KarkkainenNavarroUkkonen2000,NavarroKidaetal2001,BilleFagerbergGoertz2007},
which can be solved in time $O(n (\min \{mk, k^4 + m\}) +
\occ)$~\cite{BilleFagerbergGoertz2007}, where $n$ is the
compressed length under the LZ compression.

Theorem~\ref{thm:approxstringmatching} gives us the first non-trivial
algorithms for approximate pattern matching over any grammar
compressed string. For instance, if we plug in the Landau-Vishkin 
\cite{LV1989} or
Cole-Hariharan \cite{CH2002} algorithms in
Theorem~\ref{thm:approxstringmatching}(i) we obtain an algorithm with
$O(n(\min\{mk, k^4 + m\} + \log N) + \occ)$ time and $O(n\cdot
\alpha_k(n) +m + \occ)$ space.  Note that any
algorithm (not only the above two) and any
distance metric (not only edit distance) can be applied to
Theorem~\ref{thm:approxstringmatching}. For example, under the Hamming
distance measure we can combine our algorithm with a fast algorithm
for the (uncompressed) approximate string matching problem for the
Hamming distance measure~\cite{AmirLewensteinPorat2004}.


\subsubsection*{Tree Compression} There is a long history of tree
compression algorithms, but there appears to be little work on
rapidly navigating the compressed representation without decompresson.
In particular, 
 The DAG compression
approach has recently been applied successfully to compress XML documents \cite{BunemanCFHMV05,BusattoLM08} and
 \cite{BunemanCFHMV05} also note that
this representation aids the matching of XPath patterns, but
their algorithm partially decompresses the DAG.
Indeed \cite[p468]{BusattoLM08} specifically mention the 
problem of navigating the XML tree without decompressing the DAG,
and present algorithms whose running time is linear in the
grammar size for randomly accessing the nodes of the tree.
Jansson et al. \cite{JanssonSS12} give an ordered tree representation that supports a wide variety of navigational operations on a compressed
ordered tree.  However, their compression method is relatively 
weak---it is based solely on the degree distribution of the nodes in the tree---and cannot fully exploit repeated substructure in trees.

\subsection*{Overview} 
Before diving into technical details, we give an outline of the paper and of
the new techniques and data structures that we introduce and believe
to be of independent interest.  We first focus on the string random
acccess problem.  Let $\mathcal{S}$ be a SLP of size $n$ representing a string of length
$N$.  We begin in Section~\ref{LinearSpace} by defining a forest $H$
of size $n$ that represents the heavy paths~\cite{HarelTarjan1984} in
the parse tree of $\mathcal{S}$. We then combine the forest $H$ with
an existing \emph{weighted ancestor} data structure\footnote{A weighted
  ancestor query $(v,p)$ asks for the lowest ancestor of $v$ whose
  weighted distance from $v$ is at least $p$.}, leading to a first
solution with $O(\log N \log \log N)$ access time and linear 
space (Lemma~\ref{lem:linearspace}). The main part of the paper focuses on 
reducing the random access time to $O(\log N)$. 

In Section~\ref{IntervalBiased}, we observe that it is better to replace
the doubly-logarithmic weighted ancestor search in 
Lemma~\ref{lem:linearspace} by a (logarithmic) \emph{biased}
ancestor search.  In a biased search, we want to find the
predecessor of a given integer $p$ in a set of integers
$0 = l_0 < l_1 < \ldots < l_k = U$,
in $O(\log (U/x))$ time, where
$x = |$successor($p$) -- predecessor($p$)$|$.\footnote{Note that we need a slightly different property than so-called \emph{optimum} %
binary search trees \protect{\cite{Knuth71,Mehlhorn75}} -- we do not want to %
minimize the total external path length but rather ensure that \emph{each} %
item is at its ideal depth as in \protect{\cite{BentSleatorTarjan85}}}. 
Using biased search, the $O(\log N)$ 
predecessor queries on $H$ add up to just $O(\log N)$ time overall.
Our main technical contribution is to design two new
space-efficient data structures
that perform biased searches on sets defined by any path from a node $u \in H$
to the root of $u$'s tree. In Section~\ref{IntervalBiased}  
we describe the central
building block of the first data structure -- the {\em interval-biased
  search tree}, which is a new, simple
linear-time constructible, linear space, biased search data structure. We cannot directly use this data structure on every node-to-root path in $H$, since that would take $O(n^2)$ preprocessing time and space.
In Section~\ref{Tradeoffs} we first apply a heavy path decomposition
to $H$ itself 
and navigate between these paths using weighted
ancestor queries on a related tree $L$.  This reduces the preprocessing
time to $O(n \log n)$. To further reduce the preprocessing, we partition $L$ into disjoint
trees in the spirit of Alstrup et al.~\cite{ARTdecomposition}. 
One of these trees has $O(n/\log n)$ leaves and can be pre-processed
using the solution above. The other trees all have $O(\log n)$ leaves and 
we handle them recursively. However, before we can recurse on these trees they are
modified so that each has $O(\log n)$ vertices (rather than
leaves). This is done by another type of path decomposition (i.e. not
a heavy-path decomposition) of $L$. By carefully choosing the 
sizes of the recursive problems we get 
Theorem~\ref{thm:substringdecompression}(i) (for the case $m=1$).

For the RAM model, in Section~\ref{sec:biasedskip}, we generalize
\emph{biased skip lists} \cite{BBG}  to \emph{biased skip trees}, 
where every path from a node $u \in H$ to $u$'s root is
a biased skip list, giving the required time 
complexity. While a biased skip list takes linear 
space \cite{Iacono10}, a biased skip tree may have $\Omega(n \log N)$ pointers
and hence non-linear space, since in a biased skip list, 
``overgrown'' nodes (those with many more pointers than justified by their weight) are amortized over those ancestors which have an
appropriate number of pointers.  When used in $H$, however,
the parent of an ``overgrown'' node may have many ``overgrown'' children,
all sharing the same set of ancestors, and the amortization fails.
We note that no node will have
more than $O(\log N)$ pointers, and
use a sequence of $O(\log N)$ \emph{succinct} trees
\cite{MunroRao} of $O(|H|) = O(n)$ bits each to represent the 
skip list pointers, using $O(n \log N)$ bits
or $O(n)$ words in all.  These succinct trees 
support in $O(1)$ time a new {\em coloured ancestor query} -- a natural
operation that may find other uses -- using which 
we are able to follow skip list pointers in $O(1)$ time, giving the 
bounds of 
Theorem~\ref{thm:substringdecompression}(ii) (for the case $m=1$).

We extend both random access solutions to the substring decompression
in Section~\ref{sec:substringdecompression}, and in Section~\ref{sec:compressedapproximatestringmatching} we combine our substring decompression result with a technique of \cite{BilleFagerbergGoertz2007} to obtain an algorithm for approximate matching grammar compressed strings (giving the bounds of Theorem~\ref{thm:approxstringmatching}). 
The algorithm computes the approximate occurrences of the pattern in a single bottom-up traversal of the grammar. At each step we use the substring decompression algorithm to decode a relevant small portion of string, thus avoiding a full decompression. 

Finally, in Section~\ref{sec:labelledtrees}, we describe the differences between the random access operation in trees from that in strings.


%% file: LinearSpace.tex
\section{Fast Random Access in Linear Space}\label{LinearSpace}
In the rest of the paper, we let $\mathcal{S}$ denote an SLP of size
$n$ representing a string of length $N$, and let $T$ be the
corresponding parse tree (see Fig.~\ref{grammar}(b)). In this
section we present an $O(n)$
space representation of $\mathcal{S}$ that supports random access in
$O(\log N \log \log N)$ time, which also introduces the general
framework. To achieve this we partition 
$\mathcal{S}$ into disjoint paths according to a \emph{heavy path
  decomposition}~\cite{HarelTarjan1984}, and from these
form the \emph{heavy path forest}, which is of size $O(n)$. \\
%

\noindent
\subsubsection*{Heavy Path Decompositions}
Similar to Harel and Tarjan ~\cite{HarelTarjan1984}, we define the 
\emph{heavy path decomposition} of the parse tree $T$ as follows. For
each node $v$ define $T(v)$ to be the subtree rooted at $v$ and let
$\size(v)$ be the number of descendant leaves of $v$. We classify each
node in $T$ as either \emph{heavy} or \emph{light} based upon
$\size(v)$.\footnote{
Note that our definition of heavy paths
is slightly different than the usual one. We construct our heavy paths
according to the number of leaves of the subtrees and not the total
number nodes.} The root is
light. For each internal node $v$ we pick a child of maximum size and
classify it as heavy. The heavy child of $v$ is denoted
$\heavy(v)$. The remaining children are light. An edge to a light
child is a \emph{light edge} and an edge to a heavy child is a
\emph{heavy edge}. Removing the light edges we partition $T$ into
\emph{heavy paths}. A \emph{heavy path suffix} is a simple path $v_1,
\ldots, v_k$ from a node $v_1$ to a leaf in $T(v_1)$, such that
$v_{i+1} = \heavy(v_{i})$, for $i = 1, \ldots, k-1$.
If $u$ is a light child of $v$ then $\size(u) \leq \size(v)/2$ since
otherwise $u$ would be heavy. Consequently, the number of light edges
on a path from the root to a leaf is at most $O(\log
N)$~\cite{HarelTarjan1984}. 

We extend heavy path decomposition of trees to SLPs in a
straightforward manner. We consider each grammar variable $v$ as a
node in the directed acyclic graph defined by the grammar (see
Fig.~\ref{grammar}(c)). For a node $v$ in $\mathcal{S}$ let $S(v)$ be
the substring induced by the parse tree rooted at $v$ and define the
size of $v$ to be the length of $S(v)$. We define the heavy paths in
$\mathcal{S}$ as in $T$ from the size of each node. Since the size of
a node $v$ in $\mathcal{S}$ is the number of leaves in $T(v)$ the
heavy paths are well-defined and we may reuse all of the terminology
for trees on SLPs.  In a single $O(n)$ time bottom-up traversal of
$\mathcal{S}$ we can compute the sizes of all nodes and hence the
heavy path decomposition of $\mathcal{S}$.

\subsubsection*{Fast Random Access in Linear Space}
\label{sec:FRAiLS}
Our data structure represents
the following information for each heavy path suffix $v_1, \ldots,
v_k$ in $\mathcal{S}$.
\begin{itemize}
\item The length $\size(v_1)$ of the string $S(v_1)$.
\item The index $z$ of $v_k$ in the left-to-right order of the leaves
  in $T(v_1)$ and the character $S(v_1)[z]$.
\item A predecessor data structure for the \emph{left size sequence}
  $l_0, l_1, \ldots, l_{k}$, where $l_i$ is the sum of 1 plus the sizes of
  the left and light children of the first $i$ nodes in the heavy path
  suffix. 
\item A predecessor data structure for the \emph{right size sequence}
  $r_0, \ldots, r_{k}$, where $r_i$ is the sum of 1 plus the sizes of
  the right and light children of the first $i$ nodes in the heavy path
  suffix. 
\end{itemize}

\begin{figure}[t]
\begin{center}
\parbox{6cm}{\includegraphics[scale=0.57]{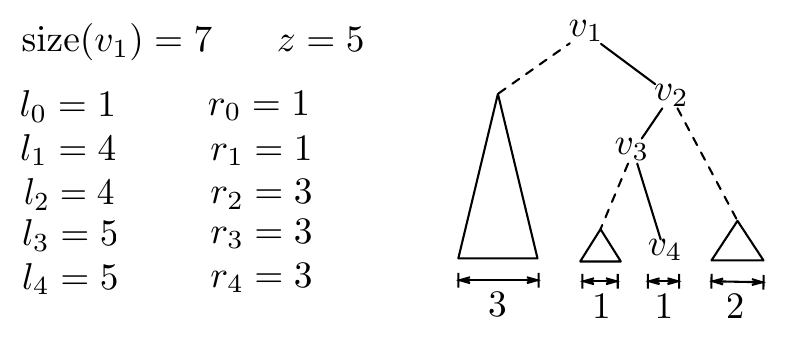}}
\parbox{1.0cm}{\ \ \ }
\caption{Ancestor search in $H$. The left and right size sequences for a heavy path suffix %
  $v_1, v_2, v_3, v_4$. The dotted edges are to light subtrees and the
  numbers in the bottom are subtree sizes. A search for $p = 5$
  returns the stored character for $S(v_1)[z]$. A search for $p = 4$
  computes the predecessor $l_2$ of $4$ in the left size
  sequence. The search continues in the left subtree of $v_3$ for
  position $p - l_2 + 1 = 4 - 4 + 1 = 1$. A search for $p = 6$ computes the
  predecessor $r_1$ of $7 - 6 = 1$ in the right size sequence. The
  search continues in the right subtree of $v_2$ for position $p - z =
  6 - 5 = 1$.}
\label{fig:searchexample}
\end{center}
\end{figure}

With this information we perform a top down search of $T$ as
follows. Suppose that we have reached node $v_1$ with heavy path
suffix $v_1, \ldots, v_k$ and our goal is to access the character
$S(v_1)[p]$. We then compare $p$ with the index $z$ of $v_k$. There
are three cases (see Fig.~\ref{fig:searchexample} for an example): 

\begin{enumerate}
\item If $p = z$  we report the stored character $S(v_1)[z]$ and end the
  search.
\item If $p < z$ we compute the predecessor $l_i$ of $p$ in the left
  size sequence. We continue the top down search from the left child $u$
  of $v_{i+1}$. The position of $p$ in $T(u)$ is $p - l_i + 1$.
\item If $p > z$ we compute the predecessor $r_i$ of $\size(v_1) - p$ in the right
  size sequence. We continue the top down search from the right child
  $u$ of $v_{i+1}$. The position of $p$ in $T(u)$ is $p-(z +
  \sum_{j = i+2}^k \size(v_j))$  (note that we can compute the sum in
  constant time as $r_k - r_{i+2}$). 
\end{enumerate}

%

The total length of all heavy path suffixes is $O(n^2)$,
thus making it unattractive to treat each suffix independently.
We show how to compactly represent all of the predecessor data
structures from the algorithm of the previous section in $O(n)$
space, and introduce the \emph{heavy path suffix forest} $H$ of
$\mathcal{S}$. The nodes of $H$ are the nodes of $S$ and a node $u$ is
the parent of $v$ in $H$ iff $u$ is the heavy child of $v$ in
$\mathcal{S}$. Thus, a heavy path suffix $v_1, \ldots, v_k$ in
$\mathcal{S}$ is a sequence of ancestors from $v_1$ in $H$. We label
the edge from $v$ to its parent $u$ by a left weight and right weight
defined as follows. If $u$ is the left child of $v$ in $\mathcal{S}$
the left weight is $0$ and the right weight is $\size(v')$ where $v'$
is the right child of $v$. Otherwise, the right weight is $0$ and the
left weight is $\size(v')$ where $v'$ is the left child of $v$.  Heavy
path suffixes in $\mathcal{S}$ consist of unique nodes and therefore
$H$ is a forest. A heavy path suffix in $\mathcal{S}$ ends at one of
$|\Sigma|$ leaves in $\mathcal{S}$ and therefore $H$ consists of
$|\Sigma|$ trees each rooted at a unique character of $\Sigma$. The
total size of $H$ is $O(n)$ and we may easily compute it from the
heavy path decomposition of $\mathcal{S}$ in $O(n)$ time.

A predecessor query on a left size sequence and right size sequence of
a heavy path suffix $v_1, \ldots, v_k$ is now equivalent to a
\emph{weighted ancestor query} on the left weights and right weights
of $H$, respectively. Farach-Colton and
Muthukrishnan~\cite{FarachMuthukrishnan1996} showed how to support
weighted ancestor queries in $O(\log \log N)$ time after $O(n)$ space
and preprocessing time. Hence, if we plug this in to our algorithm we
obtain $O(\log N \log \log N)$ query time with $O(n)$ preprocessing
time and space. In summary, we have the following result.
\begin{lemma}\label{lem:linearspace}
For an SLP $\mathcal{S}$ of size $n$ representing a string of length
$N$ we can support random access in time $O(\log N\log \log N)$ after 
$O(n)$ preprocessing time and space.
\end{lemma}


%% file: IntervalBiased.tex
\newcommand{\n}{\hat n}
\newcommand{\N}{\hat N}

\section{Interval-Biased Search Trees}\label{IntervalBiased}

In this section we reduce the 
$O(\log N \log \log N)$ random access 
time on an SLP $\mathcal{S}$ in Lemma~\ref{lem:linearspace} to 
$O(\log N)$. Recall that $O(\log N \log \log N)$ was a result of 
performing $O(\log N)$ predecessor($p$) queries, each in $O(\log \log N)$ time. 
In this section, we introduce a new predecessor data structure  
-- the {\em interval-biased search tree}.  Each predecessor($p$) query on this data structure requires $O(\log \frac U x)$ time, where $x =$ successor($p$) -- predecessor($p$), and $U$ is the universe.

To see the advantage of $O(\log \frac U x)$ predecessor queries over $O(\log \log N)$, suppose that after performing the predecessor query on the first heavy path of $T$ we discover that the next heavy path to search is the heavy path suffix originating in node $u$. This means that the first predecessor query takes $O(\log \frac {N}{|S(u)|})$ time. Furthermore, the elements in $u$'s left size sequence (or right size sequence) are all from a universe $\{0,1, \ldots, |S(u)|\}$. Therefore, the second predecessor query  takes $O(\log \frac {|S(u)|}{x})$ where $x=|S(u')|$ for some node $u'$ in $T(u)$. The first two predecessor queries thus require time  $O(\log \frac {N}{|S(u)|}+\log \frac {|S(u)|}{x}) = O(\log \frac {N}{x})$. The time required for all $O(\log N)$ predecessor queries telescopes similarly for a total of $O(\log N)$.

We next show how to construct an interval-biased search tree in linear time and space. Simply using this tree on each heavy path suffix of $\mathcal{S}$ already results in the following lemma.

\begin{lemma}
For an SLP $\mathcal{S}$ of size $n$ representing a string of length
$N$ we can support random access in time $O(\log N)$ after 
$O(n^2)$ preprocessing time and space.
\end{lemma}

\subsubsection*{A Description of the Tree}
We now define the interval-biased search tree associated with $\n$ integers $l_1 \le \ldots \le l_{\n}$ from a universe $\{0,1, \ldots, \N\}$. For simplicity, we add the elements $l_0=0$ and $l_{\n+1}=\N$. 
The interval-biased search tree is a binary tree that stores the intervals $[l_0,l_1],[l_1,l_2],\ldots, [l_{\n},l_{\n+1}]$ with a single interval in each node. The tree is described recursively:
\begin{enumerate}
\item Let $i$ be such that $(l_{\n+1}-l_0)/2 \in [l_i,l_{i+1}]$. The root of the tree stores the interval $[l_i,l_{i+1}]$.
\item The left child of the root is the interval-biased search tree storing the intervals $[l_0,l_1],\ldots, [l_{i-1},l_i]$, and the right child is the interval-biased search tree storing the intervals $[l_{i+1},l_{i+2}],\ldots, [l_{\n},l_{\n+1}]$.
\end{enumerate}


When we search the tree for a query $p$ and reach a node corresponding to the interval  $[l_i,l_{i+1}]$, we compare $p$ with $l_i$ and $l_{i+1}$. If $l_i \le p \le l_{i+1}$ then we return  $l_i$ as the predecessor. If $p < l_i$ (resp. $p >  l_{i+1}$) we continue the search in the left child (resp. right child).
Notice that an interval $[l_i,l_{i+1}]$ of length $x=l_{i+1}-l_i$ such that $\N/2^{j-1} \le x \le \N/2^j$  is stored in a node of depth at most $j$. Therefore, a query $p$ whose predecessor is $l_i$ (and whose successor is $l_{i+1}$) terminates at a node of depth at most $j$. The query time is thus $j \le 1+ \log \frac \N x =O(\log \frac \N x)$ which is exactly what we desire as $x =$ successor($p$) -- predecessor($p$). We now give an $O(\n)$ time and space algorithm for constructing the tree.

\subsubsection*{A Linear-Time Construction of the Tree}

We describe an $O(\n)$ time and space top-down construction of the interval-biased search tree storing the intervals $[l_j,l_{j+1}],\ldots, [l_k,l_{k+1}]$. We
focus on finding the interval $[l_i,l_{i+1}]$ to be stored in its root. The rest of the tree is constructed recursively so that the left child is a tree storing the intervals $[l_j,l_{j+1}], \ldots, [l_{i-1},l_i]$ and the right child is a tree storing the intervals  $[l_{i+1},l_{i+2}],\ldots, [l_k,l_{k+1}]$. 

We are looking for an interval  $[l_i,l_{i+1}]$ such that $i$ is the largest value where $l_i \le (l_{k+1}+l_j)/2$ holds. 
We can find this interval in  $O(\log(k-j))$ time by doing a binary search for $(l_{k+1}+l_j)/2$ in the subarray $l_j,l_{j+1},\ldots, l_{k+1}$. However, notice that we are not guaranteed that  $[l_i,l_{i+1}]$ partitions the intervals in the middle. In other words, $i-j$ can be much larger than $k-i$ and vice versa. This means that the total time complexity of all the binary searches we do while constructing the entire tree can amount to $O(n\log n)$ and we want $O(n)$. To overcome this, notice that 
we can find  $[l_i,l_{i+1}]$ in $\min \{ \log(i-j),\log(k-i)\}$ time if we use a  doubling search from both sides of the subarray. That is, 
if prior to the binary search, we narrow the search space by doing a parallel scan of the elements $l_j,l_{j+2},l_{j+4},l_{j+8},\ldots$ and $l_{k},l_{k-2},l_{k-4},l_{k-8},\ldots$. This turns out to be crucial for achieving $O(n)$ total construction time as we now show.

To verify the total construction time, we need to bound the total time required for all the binary searches. Let $T(\n)$ denote the time complexity of all the binary searches, then $T(\n) = T(i)+T(\n-i)+ \min \{  \log i,\log(\n-i)\}$ for some $i$. Setting $d=\min \{i,\n-i\}\le \n/2$ we get that $T(\n) = T(d) + T(\n-d) + \log d$ for some $d\le \n/2$, which is equal\footnote{By an inductive assumption that $T(\n) < 2\n - \log \n -2$ we get that $T(\n)$ is at most  $2d - \log d -2 + 2(\n-d) - \log(\n-d) -2 + \log d = 2\n-\log(\n-d)-4$, which is at most $2\n-\log \n -3$ since $d \le \n/2$.} to $O(\n)$.

%
%
%
\subsection*{Final Tuning}
We need 
one last important property of the interval-biased search tree\footnote{In fact, there exist linear-time-constructable predecessor data structures with query complexity only $O(\log \log \frac \N x)$
~\cite{MihaiPrivate}. They are more complicated than our tree, but more importantly, their query time cannot handle $\N$ reducing to $\N -l_k$.
}. Suppose that right before doing a predecessor($p$) query we know that $p>l_k$ for some $k$. We can reduce the query time to $O(\log \frac {\N -l_k} x)$ by computing for each node its lowest common ancestor with the node $[l_{\n},l_{\n+1}]$, in a single traversal of the tree. Then, when searching for $p$, we can start the search in the lowest common ancestor of $[l_k,l_{k+1}]$ and $[l_{\n},l_{\n+1}]$ in the interval-biased search tree. 


%% file: Tradeoffs.tex
\section{Closing the Time-Space Tradeoffs for Random Access}\label{Tradeoffs}

In this section we will use the interval-biased search tree to achieve $O(\log N)$ random access time but near-linear space usage and preprocessing time (instead of $O(n^2)$ as in Lemma~\ref{lem:linearspace}). 
We design a novel \emph{weighted ancestor} data structure on $H$ via a heavy path decomposition of $H$ itself. We use interval-biased search trees for each heavy path $P$ in this decomposition: one each for the left and right size sequences. It is easy to see that the total size of all these  interval-biased search trees is $O(n)$. We focus on queries of the left size sequence, the right size sequence is handled similarly.

\begin{figure*}[t]
\begin{center}
\begin{minipage}{8in}
\ \ \ \ \ \  \  \includegraphics[scale=0.4]{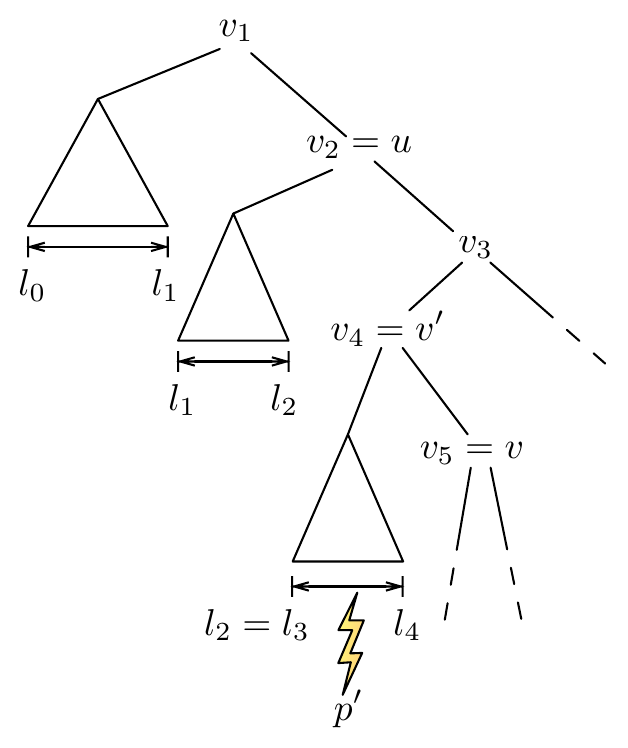} \ \ \ \ \ \ \ \ \
\includegraphics[scale=0.4]{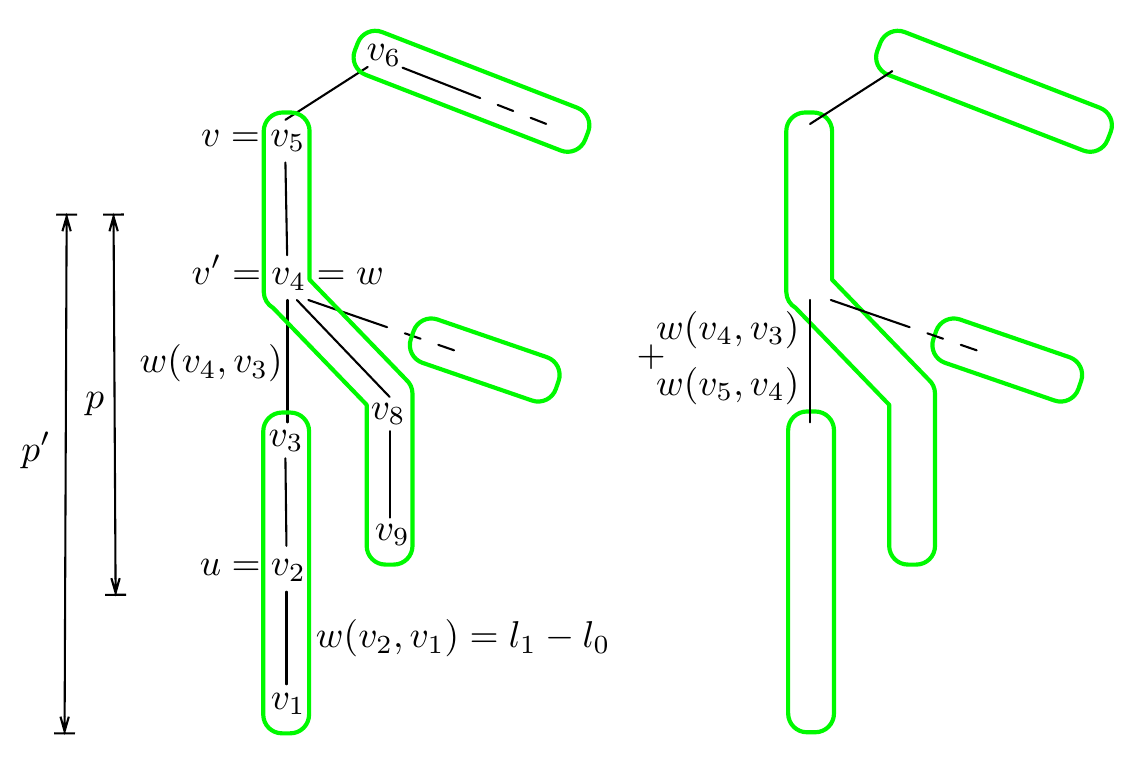}
\end{minipage}

\end{center}

\caption{The parse tree $T$ of an SLP (left), the \emph{heavy path suffix forest} $H$ (middle), and the \emph{light representation} $L$ of $H$ (right). The heavy path decomposition of $H$ is marked (in green) and defines the vertex set  of $L$.}
\label{fig:bellman}\label{THL}
\end{figure*}

Let $P$ be a heavy path in the decomposition, let $v$ be a vertex on this path, and let $w(v,v')$ be the weight of the edge between $v$ and his child $v'$, We denote by
$b(v)$ the weight of the part of $P$ below $v$ and by $t(v)$  the weight above $v$. As an example, consider the green heavy path $P = (v_5$-$v_4$-$v_8$-$v_9$) in Fig.~\ref{THL}, then $b(v_4) = w(v_4,v_8)+w(v_8,v_9)$ and $t(v_4) = w(v_5,v_4)$. In general, if $P=(v_k$-$v_{k-1}$-$\cdots$-$v_1$) then $v_1$ is a leaf in $H$ and $b(v_{i+1})$ is the $i$'th element in $P$'s predecessor data structure.  The $b(\cdot)$ and $t(\cdot)$ values of all vertices can easily be computed in $O(n)$ time.

Recall that given any vertex $u$ in $H$ and any $0\le p \le N$ we need to  be able to find the lowest ancestor $v$ of $u$ whose weighted distance from $u$ is at least $p$. If we want the total random access time to be $O(\log N)$ then finding $v$ should be done in $O\big(\log \frac {|S(u)|}{w(v,v')}\big)$ time where $v'$ is the child of $v$ which is also an ancestor of $u$.  
If both $u$ and $v$ are on the same heavy path $P$ in the decomposition, a single predecessor($p'$) query on $P$ would indeed find $v$ in $O(\log \frac {t(u)}{w(v,v')}) = O\big(\log \frac {|S(u)|}{w(v,v')}\big)$ time, where $p' = p+b(u)$. This follows from the property we described at the end of  Section~\ref{IntervalBiased}.

The problem is thus to locate $v$ when, in the decomposition of $H$, $v$ is on the heavy path $P$ but $u$ is not. To do so, we first locate a vertex $w$ that is both an ancestor of $u$ and belongs to $P$. Once $w$ is found, if its weighted distance from $u$ is greater than $p$ then $v=w$. Otherwise, a single predecessor($p''$) query on $P$ finds $v$ in $O(\log \frac {t(w)}{w(v,v')})$ time, which is $O\big(\log \frac {|S(u)|}{w(v,v')}\big)$ since $t(w) \le |S(u)|$. Here, $p'' = p$  -  weight(path from $u$ to $w$ in $H$) + $b(w)$.  We are therefore only left with the problem of finding $w$ and  the weight of the path from $u$ to $w$.

\subsubsection*{A Light Representation of Heavy paths}\label{SectionOnL}
In order to navigate from $u$ up to $w$ we introduce the \emph{light representation} $L$ of $H$. Intuitively, $L$ is a (non-binary) tree that captures the light edges in the heavy-path decomposition of $H$. Every path $P$ in the decomposition of $H$ corresponds to a single vertex $P$ in $L$, and every light edge in the decomposition of $H$ corresponds to an edge in $L$. If a light edge $e$ in $H$ connects a vertex $w$ with its child then the weight of the corresponding edge in $L$ is the original weight of $e$ plus $t(w)$. (See the edge of weight $w(v_4,v_3)+w(v_5,v_4)$ in Fig.~\ref{THL}).

The problem of locating $w$ in $H$ now translates to a weighted ancestor query on $L$. Indeed, if $u$ belongs to a heavy-path $P'$ then $P'$ is also a vertex in $L$ and locating $w$ translates to finding the lowest ancestor of $P'$ in $L$ whose weighted distance from $P'$ is at least $p-t(u)$.
As a weighted ancestor data structure on $L$ would be too costly, we utilize the important fact that the height  of $L$ is only $O(\log n)$ -- the edges of $L$ correspond to light edges of $H$ -- and construct, for every root-to-leaf path in $L$, an interval-biased search tree as its predecessor data structure. The total time and space for constructing these data structures is $O(n\log n)$. A query for finding the ancestor of $P'$ in $L$ whose weighted distance from $P'$ is at least $p-t(u)$ can then be done in $O(\log \frac {|S(u)|}{t(w)})$ time. This is 
$O\big(\log \frac {|S(u)|}{w(v,v')}\big)$ as $w(v,v') \le t(w)$. We summarize this with the following lemma.

\begin{lemma}\label{nlogn}
For an SLP $\mathcal{S}$ of size $n$ representing a string of length
$N$ we can support random access in time $O(\log N)$ after 
$O(n\log n )$ preprocessing time and space.
\end{lemma}

As noted in the Introduction, the further reduction to $O(n \alpha_k(n))$
space and preprocessing time is achieved through a further decomposition
of $L$. Intuitively, we partition $L$ into disjoint
trees in the spirit of Alstrup et al.~\cite{ARTdecomposition}. 
One of these trees has $O(n/\log n)$ leaves and can be pre-processed
using the solution above. The other trees all have $O(\log n)$ leaves and 
we want to handle them recursively. However, for the recursion to work we will need to modify these trees so that each has $O(\log n)$ vertices (rather than
leaves). As described in the following subsection, this is done by another type of path decomposition -- a \emph{branching decomposition}.

\subsection*{An Inverse-Ackerman Type bound}

We have just seen that after $O(n\log n )$ preprocessing we can support  random access in  $O(\log N)$ time. This superlinear preprocessing originates in the  $O(n\log n )$-sized data structure that we construct on $L$ for  $O\big(\log \frac {|S(u)|}{w(v,v')}\big)$-time weighted ancestor queries.
We now turn to reducing the preprocessing to be arbitrarily close to linear by recursively shrinking the size of this weighted ancestor data structure on $L$.

In order to do so, we perform a decomposition of $L$ that was originally introduced by Alstrup, Husfeldt, and Rauhe~\cite{ARTdecomposition} for solving the the \emph{marked ancestor} problem: Given the rooted tree $L$ of $n$ nodes, for every maximally high node whose subtree contains no more than $\log n$ \emph{leaves}, we designate the subtree rooted at this node a \emph{bottom tree}. Nodes not in a bottom tree make up the \emph{top tree}. It is easy to show that the top tree has at most ${n}/{\log n}$ leaves and that this decomposition can be done in linear time. 

Notice that we can afford to construct, for every root-to-leaf path  \emph{in the top tree}, an interval-biased search tree as its predecessor data structure. 
This is because there will be only ${n}/{\log n}$ such data structures and each is of size height($L$) = $O(\log n)$. In this way, a weighted ancestor query that originates in a top tree node takes $O\big(\log \frac {|S(u)|}{w(v,v')}\big)$ time as required. The problem is therefore handling queries originating in bottom trees. 

To handle such queries, we would like to recursively apply our $O(n\log n )$ weighted ancestor data structure on each one of the bottom trees. This would work nicely if the number of \emph{nodes} in a bottom tree was $O(\log n)$. Unfortunately, we only know this about the number of its \emph{leaves}. We therefore use a \emph{branching representation} $B$ for each bottom tree. The number of \emph{nodes} in the representation $B$ is indeed  $\log n$ and it is defined as follows.
 
\begin{figure}[h!]
\begin{center}
\parbox{0.6cm}{\ \    }
\parbox{5.5cm}{\includegraphics[scale=0.5]{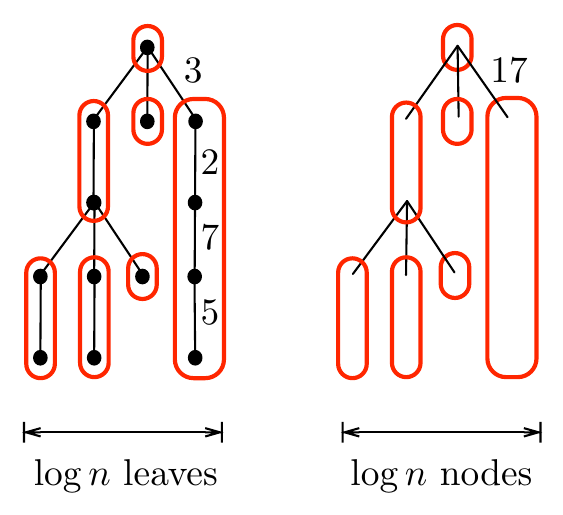}}

\caption{ A bottom tree and its branching representation $B$.  On the left is some \emph{bottom tree} -- a weighted tree with $\log n$ leaves. The bottom tree can be decomposed into $\log n$ paths (marked in red) each with at most one branching node. Replacing each such path with a single node we get the \emph{branching representation} $B$ as depicted on the right. The edge-weight 17 is obtained by the original weight 3 plus the weighted path 2+7+5. 
}
\label{B}
\end{center}
\end{figure}

We partition a bottom tree into disjoint paths according to the following rule: A node $v$ belongs to the same path as its child unless $v$ is a branching-node (has more than one child). We associate each path $P$ in this decomposition with a unique interval-biased search tree as its predecessor's data structure. The \emph{branching representation} $B$ is defined as follows. Every path $P$ corresponds to a single node in $B$. An edge $e$ connecting path $P'$ with its parent-path $P$ corresponds to an edge in $B$ whose weight is $e$'s original weight plus the total weighted length of the path $P'$ (See Fig.~\ref{B}).

Each internal node in $B$ has at least two children and therefore the number of nodes in $B$ is $O(\log n)$. Furthermore, similarly to Section~\ref{SectionOnL}, our only remaining problem is weighted ancestor queries on $B$. Once the correct node is found in $B$, we can query the interval-biased search tree of its corresponding path in $L$ in  $O\big(\log \frac {|S(u)|}{w(v,v')}\big)$ time as required. 

Now that we can capture a bottom tree with its branching representation $B$ of logarithmic size, we could simply use our 
 $O(n\log n )$ weighted ancestor data structure on every $B$. This would require an $O(\log n \log \log n)$-time construction for each one of the ${n}/{\log n}$ bottom trees for a total of $O(n \log \log n)$ construction time. 
In addition, every bottom tree node $v$ stores its weighted distance $d(v)$ from the root of its bottom tree. 
After this preprocessing, upon query $v$, we first check $d(v)$ to see whether the target node is in the bottom tree or the top tree. Then, a single predecessor query on the (bottom or top) tree 
takes $O\big(\log \frac {|S(u)|}{w(v,v')}\big)$ time as required. 

It follows that we can now support random access on an SLP in time $O(\log N)$ after only
$O(n\log \log n )$ preprocessing. In a similar manner we can use this $O(n\log \log n )$ preprocessing recursively on every $B$ to obtain an $O(n\log \log \log n )$ solution. Consequently, we can reduce the preprocessing to $O(n\log^*n )$ while maintaining  $O(\log N)$ random access. Notice that if we do this naively then the query time increases by a $\log^*n$ factor due to the $\log^*n$ $d(v)$ values we have to check. To avoid this, we simply use an interval-biased search tree for every root-to-leaf path of  $\log^*n$ $d(v)$ values. This only requires an additional $O(n\log^*n )$ preprocessing and the entire query remains $O\big(\log \frac {|S(u)|}{w(v,v')}\big)$.

Finally, we note that choosing the recursive sizes more carefully (in the spirit of~\cite{AlonSchieber87,CR1989}) can reduce the $\log^*n$ factor down to  
$\alpha_k(n)$ for any fixed $k$. This gives Theorem~\ref{ackerman}: 
\begin{theorem}\label{ackerman}
For an SLP $\mathcal{S}$ of size $n$ representing a string of length
$N$ we can support random access in time $O(\log N)$ after 
$O(n\cdot \alpha_k(n) )$ preprocessing time and space for any fixed $k$ on the pointer machine model.
\end{theorem}

%% file: biasedskip.tex
\section{Biased Skip Trees}
\label{sec:biasedskip}

In this section we give an alternate representation of
the heavy path forest $H$, that supports the ``biased''
predecessor search of the biased interval search tree;
the space and preprocessing are both $O(n)$, but 
the data structure uses the more powerful word
RAM model with word size $O(\log N)$ bits. For convenience of
description, the predecessor search is expressed a little
differently: suppose that we aim to access the $p$-th
symbol of $S(v)$ for some node $v$,
and suppose that $u$ is an ancestor of $v$ in $H$ (i.e.
$u$ is a heavy descendant of $v$ in the parse tree);
assume as previously that the desired symbol is
not the symbol associated with the root of the tree
in which $v$ is. We say that a \emph{test} at $u$
is ``true" if the desired symbol is in $u$'s heavy
child, and ``false'' otherwise; this test is perfomed
in $O(1)$ time by storing $l$ and $r$ values as before.
 Our objective is
to find the lowest ancestor $u$ in $H$ of $v$ such that
the test at $u$ is ``false''; this search should take
$O(\log (W_v / w_u ) + 1)$ time, where
for all nodes $u \in H$, $w_u = \size(u')$, where $u'$ is the light child of
$u$, and $W_u = \size(u)$.

%
%
%
%


Our solution uses a static version of {biased skip lists}~\cite{BBG},
generalized to trees.  The initial objective is to assign a
non-negative integral \emph{color} $c_v$ to each node in $v \in H$ and
there is a (logical) uni-directional linked list that
points up the tree, such that all nodes on a leaf-to-root path
whose color is at least $c$ are linked together by a series
of \emph{color-$c$ pointers}.  We defer the implementation of
color-$c$ pointers to later, but note here only
that we can follow a pointer in $O(1)$ time.
%

The biased search starting 
at a node $v$ will proceed essentially as in 
a skip list.   Let
$c^{max}_v$ denote the maximum color of any ancestor of $v$,
and $c^{max}$ the maximum color of any node in $H$.  
The search first tests $v$ -- if the answer is ``false'' we are done,
otherwise, we set $c = c^{max}_v$, and 
the current node to $u$, and suppose $v' = nca(v,c)$.  We test at $v'$;
if the outcome is ``true'' then we set the current node to
$v'$; otherwise we check that $v'$ is not the final answer
by testing the appropriate child
of $v'$.  If $v'$ is not the final answer then we set $c = c-1$
and continue.

\begin{figure}
\centering
\includegraphics[scale=0.42]{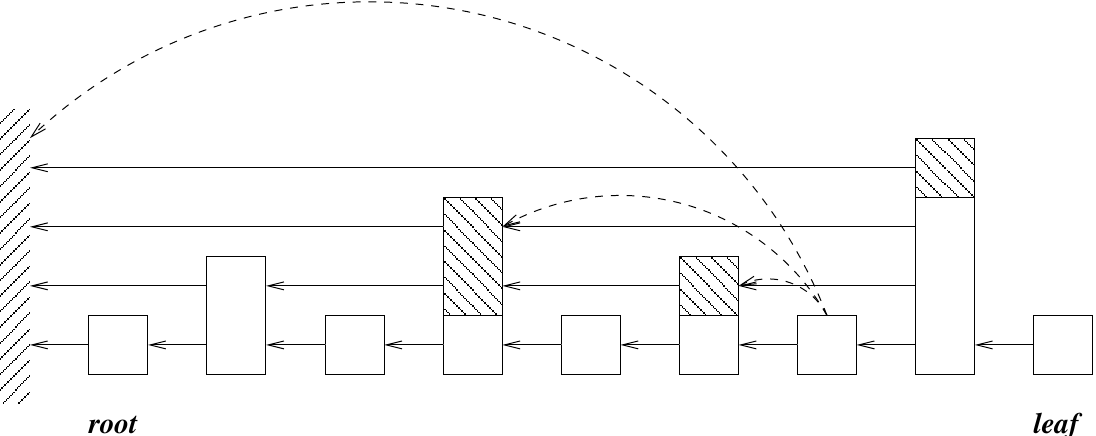}
\caption{Diagram showing the colors assigned to a sequence of vertices with ranks 1 (root), 2, 1, 1, 1, 1, 1, 3, 1 (leaf).  The %
unshaded portion of the tower of a vertex represents its rank; the shaded portion is the ``additional" pointers added by the algorithm. %
Solid pointers show explicit color-$c$ pointers that would be stored %
in a biased skip list; dotted pointers shown %
are examples of pointers %
that are available implicitly through the $nca$ operation.}
\end{figure}

We now describe how we select the colors of the nodes in $T$.
For any node $v$, denote the \emph{rank} of $v$ to be 
$r_v = \lfloor \log_2 w_v \rfloor + 1$.  We perform a pre-order traversal of each
tree in $H$. When visiting $v$,
we initially set $c_v = r_v$.  Then, while the nearest ancestor
of $v$ with color greater than or equal to $c_v$ has color 
exactly $c_v$, we increment $c_v$ by one (see Figure~1 for an example).
We now show:
\begin{lemma}
\label{lem:search}
(1) For $1 \le i \le c^{max} - 1$, between any two consecutive nodes of color $i$ there is a node of color $>i$; there is exactly one node of color $c^{max}$.
(2) $c^{max}_v \le 1 + \log_2 W_v$; $c^{max} \le 1 + \log_2 N$.
(3) For any vertex $v$ and ancestor $u$ of $v$, $c^{max}_v - c_u = O(\log (W_v/w_u))$.
\end{lemma}

\begin{proof}
(1) follows by construction. For (2) and (3), 
consider any path in $H$ from a node $v$ to the root, and as  in \cite{BBG}, 
define $N_i = | \{ u \mbox{\rm \ an ancestor of $v$} : r_u = i \} |$ and
$N'_i = | \{ u \mbox{\rm \ an ancestor of $v$} : r_u \le i \mbox{\rm \ and \ } c_u \ge i \} |$.
It is easy to see that:
\begin{equation}
N'_{i+1} \le N_{i+1} + \left \lfloor \frac{N'_i}{2} \right \rfloor
\end{equation}
From this (2) and (3) follow as in \cite{BBG}.
\end{proof}

From parts (1) and (3)
of Lemma~\ref{lem:search}, 
it follows that a search that starts at a node $v$ and ends in a node
$u$ takes $O(1+\log(W_v/w_u))$ time.  
The following lemma 
shows that one can assign colors to all the nodes in $H$ in linear time.
%
\begin{lemma}\label{lem:color-assignment}
Given $H$ and the weights of the nodes, we can compute all 
node colors in $O(n)$ time.
\end{lemma}
\begin{proof}
To assign the colors, keep a $c^{max}$-bit
counter (which fits into one word); the counter is initialized
to $0$. We perform a pre-order traversal of $H$, and when we have
visited a node $v$, the counter contains a $1$ in bit position 
$i$ (the least significant bit is position 1 and the most significant is
position $c^{max}$) if there is an ancestor of $v$ (including $v$ itself) with
color $i$, such that there is no other node with color $>i$ between
$v$ and this ancestor.  Upon arriving at a node $v$ for the first time,
we first compute $r_v$.  Taking the value of the counter at $v$'s parent
to be $x$, we set the lowest-order $r_v - 1$ bits of $x$ to 1, and add 1 to the result,
giving a value $x'$. The counter value for $v$ is in fact $x'$, and is stored with $v$.
To compute the color of $v$, we compute the bit-wise exclusive-OR of $x$ and
$x'$, and find the position of the most significant 1 bit in the result.  The implementation
of the above in constant time requires 
standard $O(1)$-time bit-wise operations, most
notably the $O(1)$-time computation of the MSB of a single word \cite{FredmanW93,Fich}. 
\end{proof}

\noindent
\subsubsection*{Nearest Colored Ancestor Problem}
We consider the following problem: Given a rooted
ordered tree $T$ with $n$ nodes, each of which 
is assigned a color from $\{1, 2, \dots, \sigma\}$, preprocess 
$T$ to answer the following query in $O(1)$ time:

\smallskip\noindent
$nca(v,c)$: given a node $v \in T$ and a color
$c$, find the lowest ancestor of $v$ in $T$ whose color
is $\ge c$.  

\smallskip
We will use this data structure for every tree in $H$; clearly, the $nca$ operation
simulates following color-$c$ pointers, thus enabling biased search.
%
%
%
%
To address our application, 
we consider the problem in the setting where word size 
$w$ is equal to the number of colors, $\sigma$. 
Our goal is to preprocess $T$ in $O(n)$ time, 
and store it in a data structure of size
$O(n)$ words (i.e., $O(n \sigma)$ bits) to support in $O(1)$-time not
only $nca()$ but also navigation queries, such as finding
the distance between an ancestor and descendant, and 
choosing the $i$-th level-ancestor of a given node.

We partition 
the string BP of length $2n$ that stores the balanced parenthesis sequence 
of the given $n$-node tree into blocks of size $b = \min\{\sigma, \lg n\}$. 
Every node in the tree belongs to either one or two different blocks.
For each block we identify a 
{\it representative} node which is the LCA of all the nodes whose corresponding 
parentheses are in that block. Thus there are $O(n/b)$ representative nodes. 
%
%
%
Our main idea is to preprocess each block so that queries whose answer lies within the 
block can be answered efficiently, as summarized in the following lemma.
In addition, in linear time we compute and store all the answers for all the representative nodes.
\begin{lemma}\label{lem:block-processing}
Given a block containing $b$ nodes where each node is associated with a color
from the range $[1,\sigma]$, one can construct a $O(n \lg \sigma)$-bit structure in 
$o(b)$ time such $nca$ queries whose  answer lies within the same block can be 
answered in constant time. 
\end{lemma}

\begin{proof}
Our first step is to reduce the set of colors within a block from $\sigma$ to $O(b)$. 
(If $\sigma = b$, this step is omitted.)
For each block, we obtain a sorted list of all colors that appear in that block. This can be
done in linear time by sorting the pairs $\langle block\_number, color_{i} \rangle$,
where $color_{i}$ is the color of the $i$-th node (i.e., the node corresponding to the
$i$-th parenthesis) in the block, using radix sort.

Let $c_{1} < c_{2} < \dots < c_{k}$, for some $k \le b$, be the set of all distinct colors 
that appear in a given block. Define $succ(c)$ to be the smallest $c_{i}$ such that 
$c_{i} \ge c$. Observe that $nca(x,c) = nca(x,succ(c))$, if the answer is within the block.
For each block, we store the sorted sequence $c_{1}, c_{2}, \dots, c_{k}$ of all 
distinct colors that appear in the block using an atomic heap~\cite{FredmanW94}, 
to support $succ()$ queries in constant time.



The range of colors in each block is now reduced to at most $b$. Thus, we need to
answer the $nca()$ query in a block of size $b$ where the nodes are associated 
with colors in the range $[1,b]$. Using $b \lg b$ bits, we store the string consisting 
of the ``reduced'' colors of the nodes, in the same order as the nodes in the block.
For each color $c$, $1 \le c \le b$, we build a $o(b)$-bit auxiliary structure that
enables us to answer the query $nca(x,c)$ in constant time, for any node $x$ in 
the block if the answer lies within the block.

We divide each block (of size $b = \lg n$) into sub-blocks of size 
$s =\epsilon \lg n/ \lg\lg n$, for some positive constant $\epsilon < 1$.
If the answer to an $nca()$ query lies in the same sub-block as the query node, then 
we can find the answer using pre-computed tables, as all the information related to
a sub-block (the parenthesis sequence and the `reduced' color information of the 
nodes) fits in $O(\lg n)$ bits -- the constant factor can be made less than $1/2$ by 
choosing the parameter $\epsilon$ in the sub-block size appropriately.
If the answer to the query does not lie in the same sub-block, but with in the same
block, then we first determine the sub-block (within the block) which contains the 
answer. To do this efficiently, we store the following additional information, for
each block.

Given a reduced color $c$  in the block and a position $i$ within the block 
(corresponding to a node $x$), we define the {\it colored excess} of the 
position (with respect to the representative of the block) 
as the number of nodes with color $c$ in the path from $x$ to $rep(x)$. 
For every reduced color in the range $[1\dots b]$ and every sub-block, 
we compute and store the minimum and maximum colored excess values within the
sub-block. Using this information for all the sub-blocks within a block, and for any 
particular color, we can find the sub-block containing the answer to a query with 
respect to that color (in constant time, using precomputed tables of negligible size). 
As there are $b/s$ sub-blocks and $b$ colors within each block, 
and the values stored for each sub-block are in the range $[0 \dots b]$, the information 
stored for each block is $O((b/s) b \lg b) = O(b (\lg\lg n)^2)$ bits. Thus, over all the 
blocks, the space used is $O(n (\lg\lg n)^2)$ bits, which is $o(n)$ words.
The computation of this information for all the  sub-blocks can be performed in $O(n)$ 
time as explained below.

The total size of the information we need to store for each sub-block is 
$O(s (\lg\lg n)^2) = O(b \lg\lg n)$ bits, and we need to be able to read the
information corresponding to all the sub-blocks within a block, corresponding to 
any particular color, by reading a constant number of $O(\lg n)$-bit ``words''.
For this, we divide the range of colors (i.e., the range $[1 \dots b]$) into 
chunks of size $d = s/\lg\lg n$, and write down the information corresponding 
to all the sub-blocks within a block, and of all the colors within a chunk, which 
fits in $O(\lg n)$ bits. Thus we can read the information  corresponding to all 
the sub-blocks within a block, corresponding to any particular color, by reading 
these $O(\lg n)$-bits.
We use precomputed tables to produce the information corresponding to
each sub-block, and for all the colors within each chunk. Hence each sub-block has 
to be ``processed'' $O(b/d)$ times (as there are $b/d$ chunks). Thus the total 
time spent producing the information for all the sub-blocks and for all the chunks 
for each block is $O((b/s) (b/d)) = O(\lg\lg n)^3$. Thus the overall time spent for 
all the blocks is $O((n/b) (\lg\lg n)^3) = o(n)$. 
\end{proof}
For each representative node $x$, we will store an array $A$ of size $\sigma$
such that $A_x[c] = nca(x,c)$, for $1 \le c \le \sigma$. As there are $O(n/b)$ 
representative nodes, and each entry in $A_x$ takes $\lg n$ bits, the total space 
used by arrays of all the representative nodes is $O((n/b) \sigma \lg n)$ bits which 
is $O(n \sigma)$. 
We will now describe how these arrays can be 
constructed with linear preprocessing time.

We first prove the following properties about the representative nodes.

\begin{lemma}
For each node x, at least one of these three statements is true: 
(i) $nca(x,c)$ lies in the (first) block to which $x$ belongs, 
(ii) $nca(x,c) = rep(x)$, or
(iii) $nca(x,c) = nca(rep(x),c)$.
\end{lemma}
\begin{proof}
The lemma follows from the following two observations:
\begin{itemize}
\item Either $nca(x,c) = parent(x)$, or $nca(x,c) = nca(parent(x),c)$.
\item $rep(x)$ is either the highest ancestor of $x$ that is within the block containing $x$,
or the lowest ancestor of $x$ that is outside the block containing $x$. (This follows from the
fact that any block that contains nodes $x$ and $y$ also contains all the nodes along the 
path between $x$ and $y$ in the tree.)
\end{itemize}
\end{proof}

\begin{lemma}
Each  representative node (except the root) has an ancestor within a height of at most 
$b$ from its level.
\end{lemma}
\begin{proof}
Consider the lowest $b-1$ ancestors of a representative node $x$. Either the 
highest node, $y$, among these which is within the same block as $x$, or $y$'s parent, 
$z$ is a representative. Note that $y$ is the LCA of all nodes between $x$ and $y$, 
and if the block contains a sibling of $y$, then $z$ is the LCA of all nodes in the block. 
\end{proof}

The root of the tree is a representative node, and the array for it consists of all null pointers.
Traverse the tree in preorder, skipping all the non-representative nodes. When a
representative node $x$ is reached, we will scan its ancestors starting from $x$ up to its 
lowest ancestor, $y$, that is also a representative. Let $A_{y}$ be the array stored at node 
$y$. During this upward scan, we will generate an array $B$ of length $\sigma$ as follows.

We keep track of the largest color value $c^{max}$ encountered at any point during 
the upward scan, and the first $c^{max}$ entries of the array $B$ are filled. In each step
of the scan, if we encounter a node whose color value is at most $c^{max}$, we simply 
skip this node. On the other hand, if we encounter a node whose color value, $c$, is 
larger than $c^{max}$, then we set the entries $B[c^{max}+1], \dots, B[c]$ to be pointers 
to the current node. We also update the value $c^{max}$ to be the new value $c$.
We now copy $A_{y}$ to another array, and overwrite the first $c_{max}$ values of $A_{y}$
with the first $c^{max}$ values of $B$. The resulting array is the array $A_{x}$ that will be 
stored at node $x$.
Generating the array $B$ takes $O(\lg n + b)$ time, as the length of $B$ is $O(\lg n)$,
and it is ``extended'' at most $b$ times. Entries of $B$ are written using bit operations on 
words (note that the word size is $\sigma$). Thus the overall running time to generate
all the arrays at the representative nodes is $O((n/b)(b+\lg n)) = O(n)$.

By plugging in this data structure in place of interval biased search trees, we get 
part(ii) of Theorems~\ref{thm:substringdecompression},
 \ref{thm:approxstringmatching} and \ref{thm:treerep}.

\section{Substring Decompression}\label{sec:substringdecompression}
We now extend our random access solutionsto
efficiently support substring decompression. Note that we can always
decompress a substring of length $m$ using $m$ random access
computations. In this section we show how to do it using just $2$
random access computations and additional $O(m)$ time. This
immediately implies Theorem~\ref{thm:substringdecompression}.

We extend the representation of $\mathcal{S}$ as follows. For each
node $v$ in $\mathcal{S}$ we add a pointer to the next descendant node
on the heavy path suffix for $v$ whose light child is to the left of
the heavy path suffix and to the right of the heavy path suffix,
respectively. This increases the space of the data structure by only a
constant factor. Furthermore, we may compute these pointers during the
construction of the heavy path decomposition of $\mathcal{S}$ without
increasing the asymptotic complexity.

We decompress a substring $S[i,j]$ of length $m = j - i$ as
follows. First, we compute the lowest common ancestor $v$ of the
search paths for $i$ and $j$ by doing a top-down search for $i$ and
$j$ in parallel. We then continue the search for $i$ and $j$
independently. Along each heavy-path on the search for $i$ we collect
all subtrees to the left of the heavy path in a linked list using the
above pointers. The concatenation of the linked list is the roots of
subtrees to left of the search path from $v$ to $i$. Similarly, we
compute the linked list of subtrees to the right of the search path
from $v$ to $j$. Finally, we decode the subtrees from the linked lists
thereby producing the string $S[i,j]$.

With our added pointers we construct the linked lists in time
proportional to the length of the lists which is $O(m)$. Decoding each
subtree uses time proportional to the size of the subtree. The total
sizes of the subtrees is $O(m)$ and therefore decoding also takes
$O(m)$ time. Adding the time for the two random access computations
for $i$ and $j$ we obtain Theorem~\ref{thm:substringdecompression}.

\section{Compressed Approximate String Matching}\label{sec:compressedapproximatestringmatching}
We now show how to efficiently solve the compressed approximate string
matching problem for grammar-compressed strings. Let $P$ and be string
of length $m$ and let $k$ be an error threshold. We assume that the
algorithms for the uncompressed problem produces the matches in sorted
(as is the case for all solution that we are aware of). Otherwise,
additional time for sorting should be included in the bounds.

To find all approximate occurrences of $P$ within $S$ without
decompressing $\mathcal{S}$ we combine our substring decompression
solution from the previous section with a technique for compressed
approximate string matching on LZ78 and LZW compressed
string~\cite{BilleFagerbergGoertz2007}.

We find the occurrences of $P$ in $S$ in a single bottom-up traversal
of $\mathcal{S}$ using an algorithm for (uncompressed) approximate
string matching as a black-box. At each node $v$ in $\mathcal{S}$ we
compute the matches of $P$ in $S(v)$. If $v$ is a leaf we decompress
the single character string $S(v)$ in constant time and run our
approximate string matching algorithm. Otherwise, suppose that $v$ has
left child $v_l$ and right child $v_r$. We have that $S(v) = S(v_l)
\cdot S(v_r)$. We decompress the substring $S'$ of $S(v)$ consisting
of the $\min\{|S(v_l)|, m+k\}$ last characters of $S(v_l)$ and the
$\min\{|S(v_r)|,m+k\}$ first characters of $S(v_r)$ and run our
approximate string matching algorithm on $P$ and $S'$. We compute the
set of matches of $P$ in $S(v)$ by merging the list of matches from
the matches of $P$ in $S(v_l)$, $S(v_r)$, $S'$ (we assume here that
our approximate string matching algorithm produces list of matches in
sorted order). This suffices since any approximate match with at most
$k$ errors starting in $S(v_l)$ and ending in $S(v_r)$ must be
contained within $S'$.

For each node $v$ in $\mathcal{S}$ we decompress a substring of length
$O(m + k) = O(m)$, solve an approximate string matching problem
between two strings of length $O(m)$, and merge lists of matches.
Since there are $n$ nodes in $\mathcal{S}$ we do $n$ substrings
decompression and approximate string matching computations on strings
of length $m$ in total. The merging is done on disjoint matches in $S$
and therefore takes $O(\occ)$ time, where $\occ$ is the total number
of matches of $P$ in $S$. With our substring decompression result from
Theorem~\ref{thm:substringdecompression} and an arbitrary approximate
string matching algorithm we obtain
Theorem~\ref{thm:approxstringmatching}.

%% file: comp-trees.tex
\section{Random Access to Compressed Trees}
\label{sec:labelledtrees}

We now consider the problem of performing operations on ``SLP-compressed'' trees.
The raw data is an ordered rooted tree $T$ (of arbitrary degree) with $N$ 
nodes.  We assume that the nodes of $T$ are numbered 
from $1$ to $N$ in pre-order, and that $T$ is represented by 
an SLP {\cal S} that generates the balanced 
parenthesis (BP) sequence of $T$  \cite{MunroRaman01}. 
As noted in the introduction, this model captures existing tree compression methods. 
We illustrate this by showing that the SLP can asymptotically match a common tree compression technique, where $T$ is compressed by sharing identical subtrees,
giving a DAG with $n$ nodes (see Fig.~\ref{fig:compressedtree}):

\begin{figure*}[t]
\begin{center}
\includegraphics[scale=0.35]{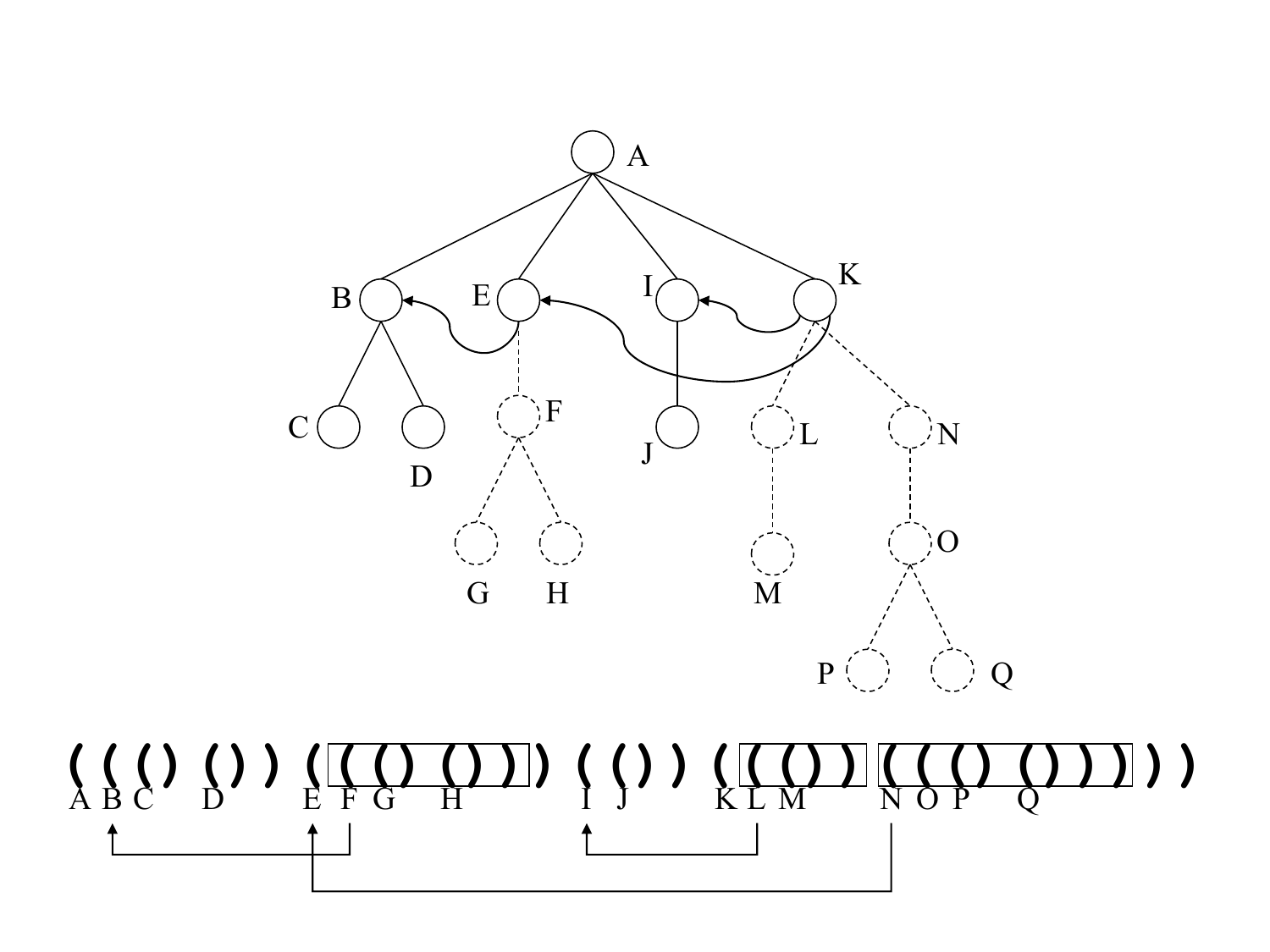}
\end{center} 
\caption{A compressed tree given as a DAG and its balanced parentheses representation.}
\label{fig:compressedtree}
\end{figure*}

\begin{figure}
\begin{center}
\includegraphics[scale=0.30]{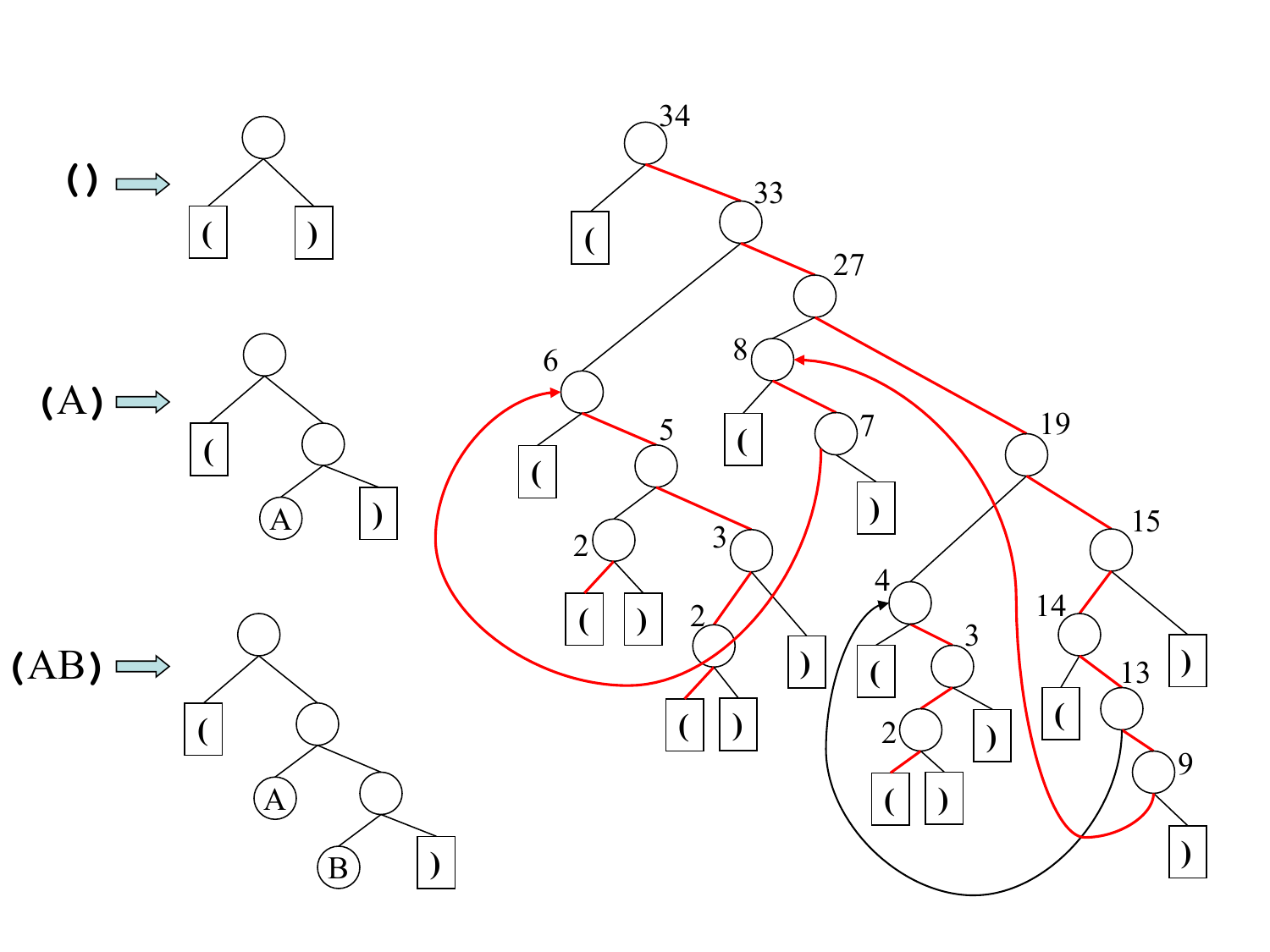} 
\caption{The SLP representing the balanced parenthesis string of the %
tree in Fig.~\ref{fig:compressedtree} -- numbers above an internal node (non-terminal) %
represent the lengths of strings output by that non-terminal.  Heavy paths %
are shown by red arrows -- there are three in all.}
\label{fig:compressedtree3}
\end{center}
\end{figure}
\begin{lemma}
Given a rooted ordered tree $T$ with $N$ nodes that is compressed to a rooted 
DAG $G$ with $e$ edges and $O(e)$ nodes, the BP string of $T$ can be
represented by an SLP of size $O(e)$.
\end{lemma}
\begin{proof}
Create an SLP that generates the  
balanced parentheses string
of the tree $T$ as follows.  For each node $x$ of the DAG $G$ with $k \ge 0$ children,
we create $k+1$ non-terminal nodes $x_0, \ldots, x_k$.  The nodes $x_0$ and $x_k$
have '(' and ')' as their left and right child, respectively.  The node $x_i$ ($0 \le i \le k-1$)
has $x_{i+1}$ as its right child, and the node $x_j$ ($1 \le j \le k$) has the corresponding
representation of the $j$-th child of the node $x$ of $G$ (see Fig.~\ref{fig:compressedtree3}).
Clearly, the size of this SLP is $O(e)$.
\end{proof}

We first consider computing some functions on a (binary) string $S$, when the string is given as an SLP ${\cal S}$. 

\begin{description}
\item[$\Rank(S, i)$:] Returns the number of 1s in $S[1]\ldots S[i]$.
\item[$\Select(S, i)$:] Returns the position of the $i$-th 1 in $S$.
\item[$\excess(S, i)$:] Returns the difference between the number of 1s and the number of 0s in
$S[1]\ldots S[i]$.
\end{description}

We omit the first argument if it is clear from the context.  In addition, we will
use $\excess$ to denote both the mathematical quantity as well as the operation
above. 
If $S$ is a balanced parenthesis string representing a tree $T$, with `(' encoded
as 1 and `)' encoded as 0, then $\excess(i)$, if the position $i$ is such that
$S[i] = \mbox{`('}$, is just the depth of the node represented by that opening
parenthesis. 

We now introduce some notation. For any binary string $s$, denote the 
number of 1s in $s$ by $\weight(s)$, and for convenience define 
$\exsum(s) = \excess(s, |s|) = 2 \cdot \weight(s) - |s|$ as the difference between the number 
of 1s and 0s in $s$.
We now show:
\begin{lemma}
\label{lem:SLPrankselect}
For an SLP ${\cal S}$ of size $n$ representing
a binary string $S$ of length $N$ we can support the operations
$\Rank$, $\Select$ and $\excess$ in $O(\log N)$ time, and $O(n \alpha_k)$ space and preprocessing 
time in the pointer machine model, and in linear space and preprocessing time on the RAM model.\end{lemma}
\begin{proof}
In what follows, we will use ${\cal T}$ to denote the parse tree of
the given string to
avoid confusion with the rooted ordered tree $T$ that we eventually
aim to represent. For any node $v \in {\cal T}$, 
abbreviate $\weight(S(v))$ and
$\exsum(S(v))$ as $\weight(v)$ and $\exsum(v)$ respectively.
If we store $\exsum(v)$ and $\weight(v)$ values at each node in 
${\cal T}$ in addition to the $\size$ values, it is 
straightforward to perform
$\Rank(i)$ and $\excess(i)$ by walking down the ${\cal T}$ 
to the $i$-th symbol 
and accumulating  weight/sum values from nodes to the left of the search path 
in $O(h)$ time, where $h$ is the height of ${\cal T}$, and $O(N)$ 
space. 

To do this in $O(\log N)$ time with $O(n)$ space, we represent the heavy path forest of ${\cal S}$ as in either
Theorem~\ref{thm:substringdecompression}(i) or (ii) and again traverse the DAG of ${\cal S}$ as though we were
accessing the $i$-th symbol of $S$. However, now with each node $v$
in the DAG  with heavy path suffix $v = v_0, v_1, \ldots, v_k$, we
store the total weight and total sum of the right and light children of $v_{1},\ldots, v_k$
(and do the same for the left and light children).  Using this information it is
easy to simulate the naive algorithm above by maintaining the invariant that after 
every biased search on a heavy path, upon exiting to a light node $v$, the accumulated
values should be the same as the values accumulated by the naive algorithm at the time
it reaches the node corresponding to $v$ in  ${\cal T}$.

It is also straightforward to perform $\Select$ in $O(h)$ time on the parse tree ${\cal T}$ by
using the weight values to guide the search to the $i$-th 1, 
and accumulating $\size$ values from the nodes to the left of the search path in
order to keep track of the position of this 1 in $S$.  In order to simulate this
in $O(\log N)$ time, we perform a new heavy-path decomposition on
${\cal S}$ using $\weight(v)$ to determine if $v$ is a heavy or light child.  In addition
we keep with each heavy path suffix, the sums of $\size$ values to the light nodes
on the right and left sides of the heavy path suffix, and simulate the naive $\Select$ algorithm using biased search on the heavy path forest
as described above.  
\end{proof}

A major tool for navigating in trees represented as a BP sequence is 
\emph{excess search}.  Specifically, the operations to be supported are:

\begin{description}
\item[$\fwd(i, \delta)$:] Given a position 
$i$ and an integer $\delta$ ($\delta$ may be positive, negative or zero), 
returns the smallest $j > i$ such that
$\excess(j) = \excess(i) + \delta$ and $-1$ if no such 
position exists. 
\item[$\bwd(i, \delta)$:] As $\fwd$, except that it returns the largest $j < i$.
\end{description}

In addition, the following operations on the BP sequence are 
useful in supporting a few additional navigational operations on the tree~\cite{SadNav10}.

\begin{description}
\item[$\minrmq(i,j)$:] Return the minimum value of $\excess(k)$, where $i \le k \le j$.
\item[$\minrmqi(i,j)$:] Return an index $k$ such that $i \le k \le j$ and $\excess(k) = \minrmq(i,j)$.
\item[$\maxrmq(i,j)$:] Return the maximum value of $\excess(k)$, where $i \le k \le j$.
\item[$\maxrmqi(i,j)$:] Return an index $k$ such that $i \le k \le j$ and $\excess(k) = \maxrmq(i,j)$.
\end{description}

We now introduce some further notation. Define $\excmax(s) = \maxrmq(1,|s|)$
as the maximum  
$\excess$ value attained at any position in $s$.  Define $\excmin(s) = \minrmq(1,|s|)$ analogously as the minimum
$\excess$ value attained at any position in $s$.  Note that $s$ need not be a binary string
representing a balanced parenthesis sequence, so  $\excmin(s)$ can be negative.
The \emph{excess range} of a string $s$ is $[\excmin(s), \excmax(s)]$.  As 
consecutive prefixes of $s$ have excess values that differ by $\pm 1$, 
every excess value within the excess range of $s$ will be achieved by some
prefix of $s$. 
We now show:
\begin{lemma}
\label{lem:SLPexcessearch}
For an SLP ${\cal S}$ of size $n$ representing
a binary string $S$ of length $N$ we can support the operations
$\fwd$, $\bwd$, $\minrmq$, $\minrmqi$, $\maxrmq$ and $\maxrmqi$, 
all in $O(\log N)$ time, and using $O(n \alpha_k)$ space and preprocessing 
time in the pointer machine model, and in linear space and preprocessing 
time on the RAM model.
\end{lemma}
\begin{proof}
The basic idea is to simulate excess search in a manner similar to the
\emph{min-max} tree \cite{SadNav10}, with the difference that the 
(logical) min-max tree is built upon the parse tree itself, and
also that excess search
in the min-max tree when it is represented as a DAG introduces some
additional challenges. 
Our description focusses on $\fwd(i, \delta)$, as $\bwd(i, \delta)$ is symmetric
(however, note that $\fwd$ may use $\bwd$ and vice-versa).  

For any node $v$ in the DAG of ${\cal S}$ with heavy path suffix $v = v_0, v_1, \ldots, v_k$, 
let $S_l(v)$ ($S_r(v)$) be the concatenation of the strings generated by the
left and light (right and light) children of $v_i$, $i = 0, \ldots, k$ 
(see Figure~\ref{fig:ex-search}).  We store the following data with $v$, in addition
to the data already stored for random access:
$\excmin(S_r(v))$, $\excmax(S_r(v))$, $\exsum(S_l(v))$ and $\exsum(S_r(v))$, abbreviated
as $\excmin_r(v), \excmax_r(v)$, $\exsum_l(v)$ and
$\exsum_r(v)$ (the asymmetry is because we focus on
$\fwd$ for now).  
Finally, suppose that $v$'s light child $u$ is a right child.  Then define
$\bar{\excmax}_r(v)$ as the maximum excess obtained within 
$S(u)$, when
$S(u)$ is considered as a substring of $S_r(v)$, i.e.
$\bar{\excmax}_r(v) = \excmax(S(u)) + \exsum_r(w)$, where
$w = v_1$ is the heavy child of $v$. If $v$'s light child is 
a left child, we take $\bar{\excmax}_r(v)$ as $-\infty$.
Define $\bar{\excmin}_r(v)$ analogously.  Create a
range maximum query  data structure \cite{HT84}
on each heavy path (if using the data structure of
Section~\ref{Tradeoffs}), or a
tree range maximum query data structure \cite{DLW09} (if using the
biased skip tree of Section~\ref{sec:biasedskip}), over
the values $\bar{\excmax}_r(v)$, and similarly
create a range minimum data structure for $\bar{\excmin}_r(v)$.
These data structures do not increase the asymptotic space 
complexity and answer any range minimum/maximum 
queries that we require in $O(1)$ time. 

\begin{figure*}[t]
\begin{center}
\includegraphics[scale=0.40]{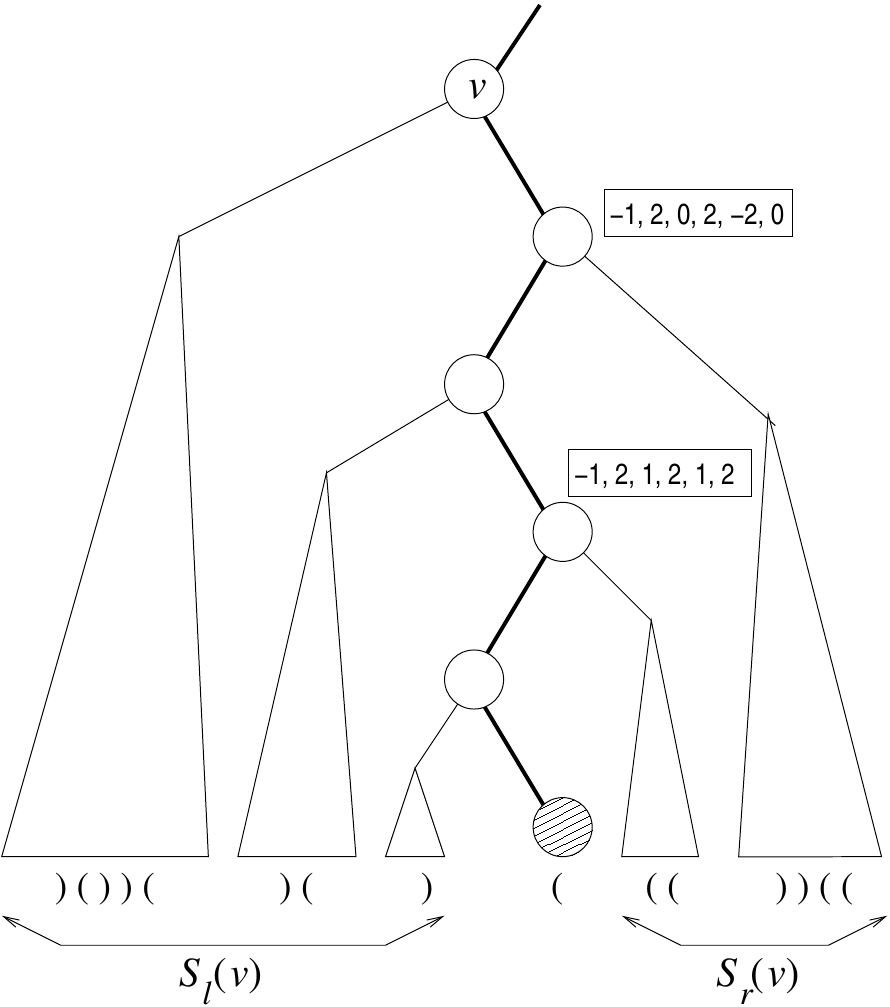}~~~~~~~~\includegraphics[scale=0.44]{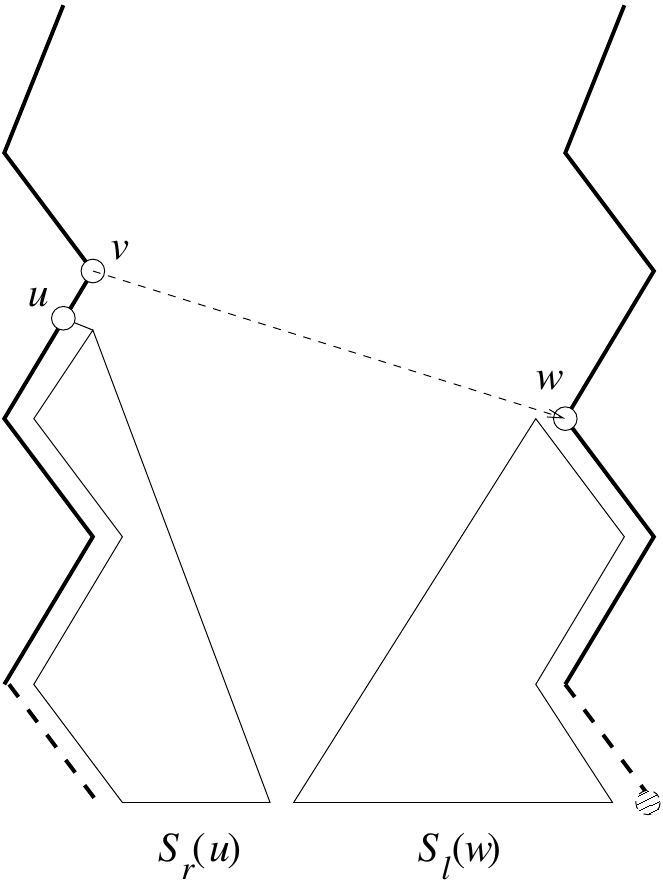}
\end{center} 
\caption{Left: $S_l(v)$ and $S_r(v)$ for a node in ${\cal T}$.  Also shown with the descendants of $v$ with right and light children are the values $\exsum_l, \exsum_r, \excmin_r, \excmax_r, \bar{\excmin}_r, \bar{\excmax}_r$. Right: Search (downward) moving from one heavy path to another.}
\label{fig:ex-search}
\end{figure*}

The operation $\fwd(i, \delta)$ is done in three phases. First, we search for the $i$-th parenthesis. 
Next, we retrace the path taken in the search backwards (in the direction of the root of ${\cal T})$, looking for the
node in ${\cal S}$ that represents the lowest common
ancestor (LCA) in ${\cal T}$ of $i$ and $j$, where
$j$ is the (unknown) position sought. Finally, we search downwards to $j$. 

The first step proceeds as previously. By construction,
this path passes through $O(\log N)$ light edges:
we record these edges.   
In the second, consider first the retracing of 
a light edge $(v, w)$, where $v$ is an ancestor of $w$, and 
assume that it has  been previously checked that 
the LCA of $i$ and $j$ is a proper ancestor of $w$.
We first check to see that $v$ is not the desired LCA,
by checking the excess range at the right heavy child of $v$
(if $v$ has no right heavy child it is anyway not the LCA).
Now suppose that $v$ is not the desired LCA and that
$(v', w')$ is the next light edge to consider.  Then
$w'$ is a heavy ancestor of $v$, and we need to check if the
LCA lies on the heavy path $w' \rightarrow v$.  This is
done by performing a range maximum and 
minimum query on the path $w' \rightarrow v$, to find the
largest value $\excmax^*$ of $\bar{\excmax}_r$ and the smallest value
$\excmin^*$ of $\bar{\excmin}_r$ achieved on this path.  If
the sought excess value does not lie in the interval 
$[\excmin^*, \excmax^*]$ then the LCA does not lie on the path
$w' \rightarrow v$ and we consider $(v', w')$ as before.
If the sought excess value lies within $[\excmin^*, \excmax^*]$ 
then the sought LCA lies on the path $w' \rightarrow v$ (the 
correctness of this argument relies on the fact that 
excess values change by $\pm 1$ per position).  Once we have
determined that the sought LCA lies on the path 
$w' \rightarrow v$, we can find the LCA using binary search
in $O(\log N)$ time using the range minimum/maximum queries
as above.  If the LCA does not lie on the path $w' \rightarrow v$
we next consider the light edge $(v', w')$.

Once the LCA $x$ is found, we move to $x$'s right light child $y$
and begin a series of biased searches along $y$'s heavy
path suffix, essentially as in the random access case.  We first 
check to see if the desired excess is achieved in $S_l(y)$, and if not,
if it is at the  non-terminal at the root of the heavy path
(if neither, it must be achieved in 
$S_r(y)$).
This check can be done by looking at the excess range of
the light child of $y$.  If the desired excess is in $S_r(y)$,
we need to find the node $z$ on the heavy
path that is closest to the root of the heavy-path tree that 
still has the desired excess in $[\excmin_r(z), \excmax_r(z)]$.
The biased search is easily adapted to this scenario, and
the desired node can be found in  $O(\log (S(x)/S(z)))$ time as
required. 

A few details need to be taken into account.  Firstly, the
excess range data that we store associated with 
$S_l(v)$ for a node $v$ are in fact based on 
\emph{backward} excesses: rather than calculating excesses
of prefixes of $S_l(v)$, we calculate excesses on suffixes
of $S_l(v)$. This is necessary so that meaningful values
can be used for $\bar{\excmin}_l(v)$ and
$\bar{\excmax}_l(v)$.  Secondly, 
we often need to adjust the target excess values appropriately.
For example, in Figure~\ref{fig:ex-search}, if the
target excess value sought in the heavy path suffix
containing $v$ was $j$, the target excess value sought
in $S(w)$ after following the light edge $(v,w)$ is
$j' = j + \exsum_r(u)$ where $u$ is $v$'s left child.  
If this target value is found not to lie in $S_l(w)$ then
the target excess value to be searched for 
in $S_r(u)$ is $j' + \exsum_l(w) \pm 1$ (depending
on whether the heavy path suffix containing
$w$ ends in a terminal labelled 0 or 1).

The operations $\minrmq(i,j)$ and $\maxrmq(i,j)$ can be supported 
in a manner similar to $\fwd$ starting from position $i$, and keeping 
track of the $m^*$ and $M^*$ values encountered so far during the 
retracing of the path, and limiting the search to within position $j$.
The operations $\minrmqi$ and $\maxrmqi$ can be immediately 
translated into $\fwd$ search once we find the $\minrmq$ and 
$\maxrmq$ values respectively (to return the leftmost indices 
satisfying the required conditions).
\end{proof}

Given an SLP ${\cal S}$ of size $n$ representing the BP sequence of a 
rooted ordered tree $T$ with $N$ nodes, we represent ${\cal S}$ using the
data structures Lemma~\ref{lem:SLPrankselect} and 
Lemma~\ref{lem:SLPexcessearch}. Now, we can support the navigational
operations on tree $T$ by using the translations of these operations to some
combinations of the operations supported by Lemma~\ref{lem:SLPrankselect} and 
Lemma~\ref{lem:SLPexcessearch} (namely, $\Rank$, $\Select$, $\excess$, $\fwd$,
$\bwd$, $\minrmq$, $\minrmqi$, $\maxrmq$ and $\maxrmqi$), as described in
\cite[Section 3]{SadNav10}.
This completes the proof of Theorem~\ref{thm:treerep}.

%% file: conclusions.tex
\section{Conclusions}

Given a string $S$ of length $N$ that is generated by a grammar of size $n$,
we have shown how to perform random access to a position in the string and
to decompress an arbitrary substring of length $m$ in time $O(\log N)$ and
$O(m + \log N)$ time respectively. We have
also shown how to perform a wide variety of operations in $O(\log N)$ time on an $N$-node 
ordered tree represented as grammar of size $n$ that generates a balanced parenthesis
string representing the tree. The data
structures have $O(n)$ space and preprocessing time on the RAM model (near-linear on
the weaker pointer machine model).  These are the first time complexities for these problems
that do not have a linear dependency on the height of the grammar.  
Using our substring decompression as a black-box, we have given the first non-trivial results on approximate string matching in grammar-compressed strings. Our black-box method is still the fastest one to date.

Recently, Verbin and Yu \cite{VerbinY13} have described 
a family of strings of size $N$, generated by a grammar of size $n$,
such that any data structure that uses $n^{O(1)}$ words of space   
must take $\Omega((\log N)^{1-\epsilon})$ time, for some constant $\epsilon > 0$,
to support random access on strings from this family. They also give
another family of strings of length $N$, generated by a grammar
of size $n = \Omega(N^{1-\epsilon})$, for some constant
$\epsilon > 0$, such that any data structure that uses $n (\log n)^{O(1)}$ words of space   
must take $\Omega(\log N / \log \log N)$ time to support random access on strings from this family. 
Both these lower bounds apply to our random access result since they are obtained on the  cell 
probe model with word size $\Theta(\log N)$, which is stronger than even the stronger of the
two models we use, and because our upper bounds are not sensitive to $n$.



%% file: SLPRAM-arXiV.bbl
\begin{thebibliography}{10}

\bibitem{AlonSchieber87}
N.~Alon and B.~Schieber.
\newblock Optimal preprocessing for answering on-line product queries.
\newblock Technical report, TR-71/87, Institute of Computer Science, Tel Aviv
  University, 1987.

\bibitem{ARTdecomposition}
S.~Alstrup, T.~Husfeldt, and T.~Rauhe.
\newblock Marked ancestor problems.
\newblock In {\em Proceedings of the 39th annual symposium on Foundations Of
  Computer Science (FOCS)}, pages 534--543, 1998.

\bibitem{AmirBensonFarach1996}
A.~Amir, G.~Benson, and M.~Farach.
\newblock Let sleeping files lie: Pattern matching in {Z}-compressed files.
\newblock {\em Journal of Comp. and Sys. Sciences}, 52(2):299--307, 1996.

\bibitem{AmirLandauSokol2003}
A.~Amir, G.~Landau, and D.~Sokol.
\newblock Inplace 2d matching in compressed images.
\newblock In {\em Proc. of the 14th annual {ACM}-{SIAM} Symposium On Discrete
  Algorithms, (SODA)}, pages 853--862, 2003.

\bibitem{AmirLewensteinPorat2004}
A.~Amir, M.~Lewenstein, and E.~Porat.
\newblock Faster algorithms for string matching with k mismatches.
\newblock {\em J. Algorithms}, 50(2):257--275, 2004.
\newblock Announced at SODA 2000.

\bibitem{ApostolicoLonardi1998}
A.~Apostolico and S.~Lonardi.
\newblock Some theory and practice of greedy off-line textual substitution.
\newblock In {\em Proc. IEEE Data compression conference}, pages 119--128,
  1998.

\bibitem{ApostolicoLonardi2000a}
A.~Apostolico and S.~Lonardi.
\newblock Compression of biological sequences by greedy off-line textual
  substitution.
\newblock In {\em Proc. IEEE Data compression conference}, pages 143--152,
  2000.

\bibitem{ApostolicoLonardi2000}
A.~Apostolico and S.~Lonardi.
\newblock Off-line compression by greedy textual substitution.
\newblock {\em Proc. IEEE}, 88(11):1733--1744, 2000.

\bibitem{ArbellLandauMitchell2001}
O.~Arbell, G.~M. Landau, and J.~Mitchell.
\newblock Edit distance of run-length encoded strings.
\newblock {\em Information Processing Letters}, 83(6):307--314, 2001.

\bibitem{BBG}
A.~Bagchi, A.~L. Buchsbaum, and M.~Goodrich.
\newblock Biased skip lists.
\newblock {\em Algorithmica}, 42:31--48, 2005.

\bibitem{BentSleatorTarjan85}
S.~W. Bent, D.~D. Sleator, and R.~E. Tarjan.
\newblock Biased search trees.
\newblock {\em SIAM J. Comput.}, 14(3):545--568, 1985.

\bibitem{BilleFagerbergGoertz2007}
P.~Bille, R.~Fagerberg, and I.~L. G{\o}rtz.
\newblock Improved approximate string matching and regular expression matching
  on ziv-lempel compressed texts.
\newblock {\em ACM Transactions on Algorithms}.
\newblock To appear. Announced at CPM 2007.

\bibitem{BunemanCFHMV05}
P.~Buneman, B.~Choi, W.~Fan, R.~Hutchison, R.~Mann, and S.~Viglas.
\newblock Vectorizing and querying large xml repositories.
\newblock In {\em ICDE}, pages 261--272. IEEE Computer Society, 2005.

\bibitem{BunkeCsirik1995}
H.~Bunke and J.~Csirik.
\newblock An improved algorithm for computing the edit distance of run length
  coded strings.
\newblock {\em Information Processing Letters}, 54:93--96, 1995.

\bibitem{BusattoLM08}
G.~Busatto, M.~Lohrey, and S.~Maneth.
\newblock Efficient memory representation of xml document trees.
\newblock {\em Inf. Syst.}, 33(4-5):456--474, 2008.

\bibitem{CegielskiGuessarianLifshitsMatiyasevich2006}
P.~C\'egielski, I.~Guessarian, Y.~Lifshits, and Y.~Matiyasevich.
\newblock Window subsequence problems for compressed texts.
\newblock In {\em Proc. of the 1st symp. on Computer Science in Russia (CSR)},
  pages 127--136, 2006.

\bibitem{Charikaretal2005}
M.~Charikar, E.~Lehman, D.~Liu, R.~Panigrahy, M.~Prabhakaran, A.~Sahai, and
  A.~Shelat.
\newblock The smallest grammar problem.
\newblock {\em IEEE Transactions on Information Theory}, 51(7):2554--2576,
  2005.
\newblock Announced at STOC 2002 and SODA 2002.

\bibitem{CR1989}
B.~Chazelle and B.~Rosenberg.
\newblock Computing partial sums in multidimensional arrays.
\newblock In {\em Proceedings of the 5th annual ACM Symposium on Computational
  Geometry (SCG)}, pages 131--139, 1989.

\bibitem{ClaudeNavarro2009}
F.~Claude and G.~Navarro.
\newblock Self-indexed text compression using straight-line programs.
\newblock In {\em Proc. 34th Mathematical Foundations of Computer Science},
  volume 5734 of {\em Lecture Notes in Computer Science}, pages 235--246, 2009.

\bibitem{CH2002}
R.~Cole and R.~Hariharan.
\newblock Approximate string matching: A simpler faster algorithm.
\newblock {\em SIAM J. Comput.}, 31(6):1761--1782, 2002.

\bibitem{CrochemoreLandauZiv-Ukelson2003}
M.~Crochemore, G.~Landau, and M.~Ziv-Ukelson.
\newblock A subquadratic sequence alignment algorithm for unrestricted scoring
  matrices.
\newblock {\em SIAM Journal on Computing}, 32:1654--1673, 2003.

\bibitem{DelprattRR08}
O.~Delpratt, R.~Raman, and N.~Rahman.
\newblock Engineering succinct dom.
\newblock In A.~Kemper, P.~Valduriez, N.~Mouaddib, J.~Teubner, M.~Bouzeghoub,
  V.~Markl, L.~Amsaleg, and I.~Manolescu, editors, {\em EDBT}, volume 261 of
  {\em ACM International Conference Proceeding Series}, pages 49--60. ACM,
  2008.

\bibitem{DLW09}
E.~D. Demaine, G.~M. Landau, and O.~Weimann.
\newblock On cartesian trees and range minimum queries.
\newblock In S.~Albers, A.~Marchetti-Spaccamela, Y.~Matias, S.~E. Nikoletseas,
  and W.~Thomas, editors, {\em ICALP (1)}, volume 5555 of {\em Lecture Notes in
  Computer Science}, pages 341--353. Springer, 2009.

\bibitem{Fich}
F.~Ellen.
\newblock Constant-time operations for words of length {$w$}.
\newblock 1999.

\bibitem{FarachMuthukrishnan1996}
M.~Farach and S.~Muthukrishnan.
\newblock Perfect hashing for strings: Formalization and algorithms.
\newblock In {\em Proceedings of the 7th Symposium on Combinatorial Pattern
  Matching (CPM)}, pages 130--140. Springer, 1996.

\bibitem{FerraginaLMM09}
P.~Ferragina, F.~Luccio, G.~Manzini, and S.~Muthukrishnan.
\newblock Compressing and indexing labeled trees, with applications.
\newblock {\em J. ACM}, 57(1), 2009.

\bibitem{FredmanW93}
M.~L. Fredman and D.~E. Willard.
\newblock Surpassing the information theoretic bound with fusion trees.
\newblock {\em J. Comput. Syst. Sci.}, 47(3):424--436, 1993.

\bibitem{FredmanW94}
M.~L. Fredman and D.~E. Willard.
\newblock Trans-dichotomous algorithms for minimum spanning trees and shortest
  paths.
\newblock {\em J. Comput. Syst. Sci.}, 48(3):533--551, 1994.

\bibitem{Gage1994}
P.~Gage.
\newblock A new algorithm for data compression.
\newblock {\em The C Users J.}, 12(2):23 -- 38, 1994.

\bibitem{Gasieniecetal2005}
L.~Gasieniec, R.~Kolpakov, I.~Potapov, and P.~Sant.
\newblock Real-time traversal in grammar-based compressed files.
\newblock In {\em Proceedings of the Data Compression Conference}, pages
  458--458, 2005.

\bibitem{GasieniecPopatov2003}
L.~Gasieniec and I.~Potapov.
\newblock Time/space efficient compressed pattern matching.
\newblock {\em Fundam. Inf.}, 56(1,2):137--154, 2003.

\bibitem{Hagerup98}
T.~Hagerup.
\newblock Sorting and searching on the word ram.
\newblock In M.~Morvan, C.~Meinel, and D.~Krob, editors, {\em STACS}, volume
  1373 of {\em Lecture Notes in Computer Science}, pages 366--398. Springer,
  1998.

\bibitem{HarelTarjan1984}
D.~Harel and R.~E. Tarjan.
\newblock Fast algorithms for finding nearest common ancestors.
\newblock {\em SIAM J. Comput.}, 13(2):338--355, 1984.

\bibitem{HT84}
D.~Harel and R.~E. Tarjan.
\newblock Fast algorithms for finding nearest common ancestors.
\newblock {\em SIAM J. Comput.}, 13(2):338--355, 1984.

\bibitem{HermelinLandauLandauWeimann2009}
D.~Hermelin, S.~Landau, G.~Landau, , and O.~Weimann.
\newblock A unified algorithm for accelerating edit-distance via
  text-compression.
\newblock In {\em Proc. of the 26th International Symposium on Theoretical
  Aspects of Computer Science (STACS)}, pages 529--540, 2009.

\bibitem{Iacono10}
J.~Iacono.
\newblock private communication.
\newblock 2010.

\bibitem{JanssonSS12}
J.~Jansson, K.~Sadakane, and W.-K. Sung.
\newblock Ultra-succinct representation of ordered trees with applications.
\newblock {\em J. Comput. Syst. Sci.}, 78(2):619--631, 2012.

\bibitem{KarkkainenNavarroUkkonen2000}
J.~Karkkainen, G.~Navarro, and E.~Ukkonen.
\newblock Approximate string matching over {Z}iv-{L}empel compressed text.
\newblock In {\em Proc. of the 11th symposium on Combinatorial Pattern Matching
  (CPM)}, pages 195--209, 2000.

\bibitem{KarkkainenUkkonen1996}
J.~Karkkainen and E.~Ukkonen.
\newblock {L}empel-{Z}iv parsing and sublinear-size index structures for string
  matching.
\newblock In {\em Proc. of the 3rd South American Workshop on String Processing
  (WSP)}, pages 141--155, 1996.

\bibitem{KiefferYang2000}
J.~C. Kieffer and E.~H. Yang.
\newblock Grammar based codes: A new class of universal lossless source codes.
\newblock {\em IEEE Trans. Inf. Theory}, 46(3):737--754, 2000.

\bibitem{Kiefferetal2000}
J.~C. Kieffer, E.~H. Yang, G.~J. Nelson, and P.~Cosman.
\newblock Universal lossless compression via multilevel pattern matching.
\newblock {\em IEEE Trans. Inf. Theory}, 46(5):1227 -- 1245, 2000.

\bibitem{Knuth71}
D.~E. Knuth.
\newblock Optimum binary search trees.
\newblock {\em Acta Informatica}, 1:14--25, 1971.

\bibitem{LV1989}
G.~M. Landau and U.~Vishkin.
\newblock Fast parallel and serial approximate string matching.
\newblock {\em J. Algorithms}, 10(2):157--169, 1989.

\bibitem{LarssonMoffat2000}
J.~N. Larsson and A.~Moffat.
\newblock Off-line dictionary-based compression.
\newblock {\em Proc. IEEE}, 88(11):1722 -- 1732, 2000.
\newblock Announced at DCC 1999.

\bibitem{Lifshits2007}
Y.~Lifshits.
\newblock Processing compressed texts: {A} tractability border.
\newblock In {\em Proc. of the 18th symposium on Combinatorial Pattern Matching
  (CPM)}, pages 228--240, 2007.

\bibitem{MakinenNavarroUkkonen1999}
V.~Makinen, G.~Navarro, and E.~Ukkonen.
\newblock Approximate matching of run-length compressed strings.
\newblock In {\em Proc. of the 12th Symposium On Combinatorial Pattern Matching
  (CPM)}, pages 1--13, 1999.

\bibitem{Manber1994}
U.~Manber.
\newblock A text compression scheme that allows fast searching directly in the
  compressed file.
\newblock In {\em Proc of the 5th Symposium On Combinatorial Pattern Matching
  (CPM)}, pages 31--49, 1994.

\bibitem{Mehlhorn75}
K.~Mehlhorn.
\newblock Nearly optimal binary search trees.
\newblock {\em Acta Informatica}, 5:287--295, 1975.

\bibitem{MunroRaman01}
J.~I. Munro and V.~Raman.
\newblock Succinct representation of balanced parentheses and static trees.
\newblock {\em SIAM J. Comput.}, 31(3):762--776, 2001.

\bibitem{MunroRao}
J.~I. Munro and S.~S. Rao.
\newblock Succinct representations of functions.
\newblock In {\em International Colloquium on Automata, Languages and
  Programming (ICALP)}, pages 1006--1015, 2004.

\bibitem{Navarro2001a}
G.~Navarro.
\newblock A guided tour to approximate string matching.
\newblock {\em ACM Comput. Surv.}, 33(1):31--88, 2001.

\bibitem{NavarroKidaetal2001}
G.~Navarro, T.~Kida, M.~Takeda, A.~Shinohara, and S.~Arikawa.
\newblock Faster approximate string matching over compressed text.
\newblock In {\em Proc. of the 11th Data Compression Conference (DCC)}, pages
  459--468, 2001.

\bibitem{Nevill-ManningWitten1997}
C.~G. Nevill-Manning and I.~H. Witten.
\newblock Identifying hierarchical strcture in sequences: A linear-time
  algorithm.
\newblock {\em J. Artif. Intell. Res. (JAIR)}, 7:67--82, 1997.

\bibitem{MihaiPrivate}
M.~P\v{a}tra\c{s}cu.
\newblock private communication.
\newblock 2009.

\bibitem{Rytter2003}
W.~Rytter.
\newblock Application of {Lempel}-{Ziv} factorization to the approximation of
  grammar-based compression.
\newblock {\em Theoretical Computer Science}, 302(1-3):211--222, 2003.

\bibitem{SadNav10}
K.~Sadakane and G.~Navarro.
\newblock {Fully-Functional Succinct Trees}.
\newblock In {\em Proc. ACM-SIAM SODA}, pages 134--149, Jan. 2010.

\bibitem{Sellers1980}
P.~Sellers.
\newblock The theory and computation of evolutionary distances: Pattern
  recognition.
\newblock {\em J. Algorithms}, 1(4):359--373, 1980.

\bibitem{Shibata-et-al-1999}
Y.~Shibata, T.~Kida, S.~Fukamachi, M.~Takeda, A.~Shinohara, T.~Shinohara, and
  S.~Arikawa.
\newblock Byte {P}air encoding: {A} text compression scheme that accelerates
  pattern matching.
\newblock {\em Technical Report DOI-TR-161, Department of Informatics, Kyushu
  University}, 1999.

\bibitem{Shibata-et-al-2000}
Y.~Shibata, T.~Kida, S.~Fukamachi, M.~Takeda, A.~Shinohara, T.~Shinohara, and
  S.~Arikawa.
\newblock Speeding up pattern matching by text compression.
\newblock In {\em Proc. of the 4th Italian Conference Algorithms and Complexity
  (CIAC)}, pages 306--315, 2000.

\bibitem{Tarjan83}
R.~E. Tarjan.
\newblock {\em Data Structures and Network Algorithms}.
\newblock SIAM, 1983.

\bibitem{VerbinY13}
E.~Verbin and W.~Yu.
\newblock Data structure lower bounds on random access to grammar-compressed
  strings.
\newblock In J.~Fischer and P.~Sanders, editors, {\em CPM}, volume 7922 of {\em
  Lecture Notes in Computer Science}, pages 247--258. Springer, 2013.

\bibitem{Welch1984}
T.~A. Welch.
\newblock A technique for high-performance data compression.
\newblock {\em IEEE Computer}, 17(6):8--19, 1984.

\bibitem{YangKieffer2000}
E.~H. Yang and J.~C. Kieffer.
\newblock Efficient universal lossless data compression algorithms based on a
  greedy sequential grammar transform -- part one: Without context models.
\newblock {\em IEEE Trans. Inf. Theory}, 46(3):755--754, 2000.

\bibitem{ZivLempel1976}
J.~Ziv and A.~Lempel.
\newblock On the complexity of finite sequences.
\newblock {\em IEEE Transactions on Information Theory}, 22(1):75--81, 1976.

\bibitem{ZivLempel1977}
J.~Ziv and A.~Lempel.
\newblock A universal algorithm for sequential data compression.
\newblock {\em IEEE Transactions on Information Theory}, 23(3):337--343, 1977.

\end{thebibliography}
